%% file: main.tex
\documentclass[conference,compsoc]{IEEEtran}
\ifCLASSOPTIONcompsoc
  \usepackage[nocompress]{cite}
\else
  \usepackage{cite}
\fi
\ifCLASSINFOpdf
\else
\fi
\usepackage{enumitem}
\usepackage{tikz}
\usepackage{tabularx}
\newcolumntype{Y}{>{\centering\arraybackslash}X}

\newcolumntype{Z}{>{\hsize=1.2\hsize}X}
\newcolumntype{Q}{>{\hsize=.8\hsize}X}
\newcolumntype{V}{>{\hsize=.15\hsize}X}
\usepackage{caption}
\usepackage{subcaption}
\usepackage{stfloats}
\usepackage{multirow}
\usepackage{multicol}
\usepackage{makecell}
\usepackage{balance}
\usepackage{hyperref}
\usepackage{nth}
\usepackage{xurl}
\usepackage{censor}
\usepackage{verbatim}
\usepackage{booktabs}
\usepackage{fontawesome5}
\usepackage{xcolor}

\begin{document}
%


\title{``Oh, sh*t! I actually opened the document!'': An Empirical Study of the Experiences with Suspicious Emails in Virtual Reality Headsets}


\author{\IEEEauthorblockN{Filipo Sharevski}
\IEEEauthorblockA{School of Computing \\
DePaul University\\
Chicago, IL 60604\\
Email: fsharevs@depaul.edu}
\and
\IEEEauthorblockN{Jennifer Vander Loop}
\IEEEauthorblockA{School of Computing \\
DePaul University\\
Chicago, IL 60604\\
Email: jvande27@depaul.edu}
\and
\IEEEauthorblockN{Sarah Ferguson}
\IEEEauthorblockA{School of Computing \\
DePaul University\\
Chicago, IL 60604\\
Email: sfergu12@depaul.edu}
}


%


\maketitle

\begin{abstract}
This paper reports on a study exploring user experiences with suspicious emails and associated warnings when accessed through virtual reality (VR) headsets in realistic settings. A group of (\textit{n}=20) Apple Vision Pro and another group of (\textit{n}=20) Meta Quest 3 users were invited to sort through their \textit{own} selection of Google mail suspicious emails through the VR headset. We asked them to verbalize the experience relative to how they assess the emails, what cues they use to determine their legitimacy, and what actions they would take for each suspicious email of their choice. We covertly sent a \textit{false positive} suspicious email containing either a URL or an attachment (an email that is assigned a suspicious email warning but, in reality, is a legitimate one) and observed how participants would interact with it. Two participants clicked on the link (Apple Vision Pro), and one participant opened the attachment (Meta Quest 3). Upon close inspection, in all three instances, the participant ``fell'' for the phish because of the VR headsets' hypersensitive clicking and lack of ergonomic precision during the routine email sorting task. These and the other participants thus offered recommendations for implementing suspicious email warnings in VR environments, considerate of the immersiveness and ergonomics of the headsets' interface.

\end{abstract}


%
\IEEEpeerreviewmaketitle

\input{sections/01.introduction}

\input{sections/02.background}

\input{sections/03.study}
\input{sections/04.results}
\input{sections/05.discussion}
\input{sections/06.conclusion}

\bibliographystyle{IEEEtran}
\bibliography{references}
%

\input{sections/appendix}

\end{document}

%% file: sections/01.introduction.tex
\section{Introduction} \label{sec:introduction}
The commercial proliferation of Virtual Reality (VR) head-mounted devices (or headsets) such as Meta Quest and Apple Vision Pro offered users the possibility to participate in various activities in more immersive ways than before~\cite{Stephenson2022}. Gaming was the go-to VR activity from the start, though activities such as social collaboration~\cite{Freeman2022}, clinical care~\cite{Fu2022}, education~\cite{Drey2022}, and shopping~\cite{Ward2023} quickly capitalized on the immersiveness. Super-imposing 3D virtual objects on the real world is an attractive mode of interaction that has already tempted 13\% of the US consumers to acquire a virtual reality headset~\cite{Atopia2023} and tempt at least 21\% of the consumer base to do so in the near future~\cite{Frankel2024}.

As more and more people expressed strong interest in VR headsets~\cite{Frankel2024}, companies like Apple and Meta are looking for domains that could accelerate the adoption of their products, such as productivity~\cite{Gonzalez-Franco2024}. The Apple Vision Pro is marketed as a productivity tool or a ``workspace with infinite space'' \cite{Apple2024} and the Quest offers the ``Infinite Office", a virtual workspace with the software needed for work and communication, including email functionality~\cite{Hayden2020}. Though still largely an experimental idea, the early adopters hailed them as ``the ultimate work-from-home devices'' that allow for ``peripersonal task organization,'' ``direct attention'' and ``full-body interaction'' experiences \cite{Hart2024, Axon2024, Miller2024}. 
 
Transferring the understanding of desktop and mobile computing into the virtual reality environment, especially for work, is non a trivial task as the immersive interaction is foreign to most people~\cite{Gonzalez-Franco2024}. The slow roll-out, so far, is focused on experiential tasks such as Zoom meetings, individual tasks such as training, maintaining interaction among physically distanced coworkers, and providing workers with private spaces in public settings~\cite{Cheng2024, Guo2019, Grubert2018}. PricewaterhouseCoopers (PwC), for example, purchased thousands of VR headsets to help employees combat Zoom fatigue and conduct soft-skills training \cite{Makortoff2021}. 

A task integral to any productivity-driven use of VR headsets is \textit{email correspondence}. Provided the aid of peripherals, sorting, reading, and responding to emails is the key element for a seamless transition to full workplace immersion, given the VR headsets' particular advantage when it comes to multitasking~\cite{Mcgill2020}. Dealing with emails, some of which are phishing or spam, is a perennial security problem affected not just by multitasking~\cite{Burda2024} but also the general pressures of the working environment~\cite{Lain2022, Distler2023}. As such, the question arises of how people would respond to suspicious emails when they access them through VR headsets. In conventional desktop and mobile computing, people do have the aid of warnings in cuing about potential spam or phishing emails~\cite{Petelka2019, Volkamer2017} so another question arises how these warnings render the necessary anti-deception friction in a virtual reality environment. 

To our knowledge, so far, no one attempted to study how people experience suspicious emails through VR headsets. We took it upon ourselves to obtain answers in realistic settings and also ask for recommendations towards a seamless VR email correspondence transition. Our research questions guiding our study, thus, were:  

\vspace{1em}

\begin{itemize}
    \itemsep 0.7em

    \item \textbf{RQ1:} How do users assess suspicious emails sent to their \textit{own} email addresses through (i) Meta Quest 3; and (ii) Apple Vision Pro VR headsets?

    \item \textbf{RQ2:} What actions do users take about suspicious emails sent to their \textit{own} email addresses through (i) Meta Quest 3; and (ii) Apple Vision Pro VR headsets?
   
    \item \textbf{RQ3:} What recommendations do users have for usability and VR immersiveness/ergonomics improvements of banner warnings about suspicious emails when accessed through VR headsets?

\end{itemize}

We obtained approval from our Institutional Review Board (IRB) to conduct a mild deception study with a sample of \textit{n}=40 participants (20 in the Apple Vision Pro and 20 in the Meta Quest 3 groups, respectively) who had a previous experience with VR headsets. As the Apple Vision Pro is prohibitively expensive to be purchased in quantities, we invited the participants to join a physical collaborative space and use this headset or the Meta Quest 3 for the purpose of studying how users utilize virtual reality headsets to interact with emails in general. We used this ``cover'' in order to situate our study in \textit{realistic} settings, that is, to allow for participants to interact with emails sent to their \textit{own} email address and \textit{own} phone device, per the methodology guidelines for outlined in~\cite{Sharevski2024-usenix}. Once they set up the VR headset and logged into their email address (we used Gmail as a preferred provider) we asked the participants to perform a few tasks with emails of their choice during, which they verbalized their steps and provided their experiences/opinions. 

Respecting the privacy of their personal email correspondence in their main inbox, the first task was to review several emails of their choice in their spam/junk folder with their assistive technology of choice and verbalize to us only the subject line and sender. Using this verbalization, we ensured that these emails were not dangerous (e.g., resembled standard spam or phishing) both by checking for pretext and formatting patterns found in databases of known spam/phishing emails~\cite{openphish} and using our own assessment experience. If we were uncertain, we asked the participant to proceed to the next one (during the debriefing, we advised them that it was best to delete it). We then asked the participants to open the email, assess the legitimacy of the email [\textbf{RQ1}] (with a baseline established beforehand, as described in Appendix \ref{app:baseline}), and share with us what the most likely action they would perform on it [\textbf{RQ2}] (without actually clicking any links or downloading any attachments). 
    
To ensure that participants encountered both phishing and spam banner warnings (which are more prevalent in the spam/junk folders than the main inbox), we secretly sent a \textit{false positive} phishing email to their own email address --- an email that providers deem as ``phishing,'' assign a phishing banner warning, and automatically put it in the spam/junk folder --- but in reality the email is legitimate. Once we completed the assessment tasks, we asked our participants about their accessibility and usability design recommendations for the banner warnings, as well as their real-life experiences with detecting and acting on phishing and spam emails in general.
Once we completed the assessment tasks, we asked our participants about their recommendations for improving the email correspondence experience, particularly the design of the suspicious email warnings [\textbf{RQ3}]. 
    
Our findings indicate that the warnings were effective at encouraging users not to interact with suspicious emails. About 75\% of participants who encountered a warning indicated that they would not interact with the email, whereas only 20\% of participants who did not encountered a warning banner elected not to interact with the email. The most commonly encountered warning was the one for spam messages, which participants felt provided them with little to no information about the safety of the email. All of the participants who encountered warnings alerting them that the email was either (i) dangerous, (ii) likely phishing, (iii) contained potentially unsafe images, (iv) or came from a blocked sender used these warnings as cues when deciding if an email was safe and what action to take from it.

Participants expressed reservations about the way the spam banner warnings were formatted (see Figure~\ref{fig:gmail-spam} as they felt it failed to help people correctly decide what action to take with an email through a VR headset. The suggestions included implementing a color coding system (instead of only gray), improving the message content to contain a more detailed description of the safety of the email (including a severity or risk level indicator), and adding pop-ups to ask users if they really want to click on unsafe elements of the email as the VR interface allows for such an immersive interaction. Having the option of using a VR headset for productivity intrigued our participants, though many found that the VR headset itself created additional challenges in determining the safety of a potential spam or phishing email. 

Participants reported they were distracted by seeing their physical surroundings and felt it would help if their focus was directed to their email better. Five, or 25\% of the Apple Vision Pro participants complained about how ``frictionless'' it was to click on a link or any button (conjecturing that this might result in an accidental phishing that, even if rectified, would result in loss of productivity). Four, or 20\% of the Meta Quest 3 participants felt that they were not able to click with the precision needed to perform a task such as inspecting a URL or an attachment by hovering over it. These interactive hurdles led one participant in each group to click on our test \textit{false positive} link and one in the Meta Quest 3 participant opened the test \textit{false positive} attachment.

\noindent \textbf{The main contributions of this paper are}: 

\begin{itemize}
\itemsep 0.7em
    \item Empirical evidence of the way individuals assess suspicious emails using their \textit{own} email addresses through VR headsets;  

    \item Indications of VR-specific phishing susceptibility factors due to hypersensitivity and lack of ergonomic precision in the interface;

    \item Design recommendations for suspicious email warnings' adaptation warnings within VR environments.

\end{itemize}

\vspace{-1em}

%% file: sections/02.background.tex
\section{Background} \label{sec:background}

\subsection{VR Threat Landscape}
VR headsets have so far been subjected to threat evaluation concerning authentication and leakage of private information. As the immersiveness is seen as the long-awaited interaction affordance that will offer an alternative to the conventional entering of credentials~\cite{Stephenson2022}, The threats to VR headsets could come from adversaries with and without access to the VR headset. An adversary without access or a ``shoulder-surfer'' might not be able to directly snoop on the credential entering process of VR users~\cite{Duzgun2022} but could craft a careful observation based attack that would enable deciphering approximately between 75-80\% of text inputs made in VR headset~\cite{Arafat2021, Gopal2023, Khalili2024, luo2024eavesdropping, Slocum2023}, coming effectively close to conventional keylogging attacks.

To overcome these observational attacks that focus on variation in the pose of the users' head and hand gestures, Apple Vision Pro relies on an eye tracker to minimize the need for any movements while authenticating or entering sensitive information. But an adversary with access, particularly one able to exploit the Apple Persona is still able to infer eye-related biometrics from the avatar image to reconstruct text entered via gaze-controlled typing \cite{wang2024gazeploit}. Or worse, an adversary with access could simply launch an immersive hijacking attack on a Meta Quest 3, or control a user’s interaction with their VR headset by trapping them inside a malicious app that masquerades as the full VR interface, effectively gaining access to all the users' private information, including credentials~\cite{yang2024inception}. 

With the accelerated adoption of VR headsets for productivity, we see an additional third type of adversary that, independently of the type of access, is able to recondition the well-known social engineering approach for stealing users' credentials or private information through the use of phishing and spamming emails. This adversary is not bound to a particular VR headset type, nor does it need any specific and complicated computational setup to be able to successfully launch an attack. Granted, biometric authentication schemes that do not depend on user input~\cite{Duzgun2022} might alleviate this threat, to an extent, though many of them might still not be integrated with traditional productivity applications and it will take some time to do so.

\subsection{Threat Vector: Suspicious Emails}
Suspicious emails contain malicious URLs and attachments aiming to steal users' credentials or install malware. Spotting a suspicious email is a delicate task that the majority of users haven't fully yet mastered as malicious URLs and attachments still get ``clicked'' in concerning numbers -- the average successful click rate for a phishing attack consistently remains close to 20\% over the years \cite{Verizon.2022, Verizon.2023}). As email communication allows for easy deception through impersonation and influencing pretexts~\cite{Burda2024}, users must rely on ``warnings'' or cues that alert them about impending phishing emails, usually generated by their email providers (if providers like Google correctly detect phishing or a spam attempt; a non-negligible amount of suspicious emails manage to get through users' inboxes undetected and equal among of legitimate emails get incorrectly assigned a suspicious email warning~\cite{Kumaran2022, Microsoft2024}).

The suspicious email warnings thus are implemented as interventions in the user interfaces that are placed \textit{in situ} while a user is engaged in an email correspondence task, i.e., reading, sorting, or responding to an email~\cite{Franz2021}. The goal is to either force the attention of a user to a suspicious element like a URL or attachment (usually within an email client)~\cite{Petelka2019} or offer options for users to ``go back to safety'' or ``proceed to a website'' (usually in a browser). This is done by placing the warnings in banner variants before the email subject lines \cite{GooglePhishing, OutlookPhishing}, displaying red question marks or phishing hooks in the graphical avatar icon of the sender~\cite{Risher-Miller.2017}, or showing just-in-time, just-in place URL trustworthiness tips in the email body~\cite{Volkamer2017}. Often, the interactive warnings come with additional informative messages that communicate the threats to ``inoculate'' users against future suspicious email correspondence.`

Evidence from user evaluations of the interactive warnings suggests that users tend to adhere to the interactive warnings~\cite{Franz2021}, provided the wording is comprehensible and the design prevents habituation~. The adherence and phishing safety, however, does not come without a cost -- usually, the forced attention is distracting, time-consuming, and tedious~\cite{Wash2020}, especially with a high number of emails a user receives a day and the fractured attention due to multitasking~\cite{Mossano2023}. There is also a difference in effectiveness whether the warning ``friction'' happens within an email client as a banner (the usual vector for delivery of phishing attacks \cite{Verizon.2023}) or in a browser as a splash screen, with the later implementation being better preventing participants from reaching phishing websites~\cite{Petelka2019}.  

\subsection{VR and Suspicious Emails}
The emergent nature of the mixed reality continuum defined by Milgram and Kishino~\cite{milgram1994taxonomy} restricts the work done so far relative to how users interact with and experience suspicious emails aiming to capitalize either on the augmentation or the virtuality for the purpose of phishing, spamming, or scamming. So far, Kanaoka and Isohara~\cite{Kanaoka2024} used the augmentation (e.g., AR glasses) to help users analyze images of URLs displayed across devices towards better phishing detection. Their usability test showed that the AR glasses helped users improve the correct identifying of phishing attempts compared to a baseline condition. Jansen and Fischbach~\cite{JansenFischbach2020} developed a ``social engineering'' educational VR game where the player has to perform a voice phishing attack to complete all missions. Bakker~\cite{Bakker2024} created a VR social engineering game to teach users about the most prevalent cues of deception implemented in phishing emails (anomalies in the senders' email address, grammar and spelling mistakes in the email body, principles of influence). A preliminary test with five students showed that all of them scored higher on the HAIS-Q after having played the game, compared to their score before.

%% file: sections/03.study.tex
\section{Study} \label{sec:study}

\subsection{Study Methodology} \label{sec:mthodology}
To determine how users interact with suspicious emails and the associated phishing and spam warnings in VR headsets, we directed our participants to access several emails within their Gmail spam folder. This would have constituted a priming to phishing/spam if we had done so at the beginning of the study, therefore, we asked them a couple of questions prior to starting efforts to frame the study in a broader suspicious email context. The first question was about how familiar they are with VR headsets, and the second was about how they prefer to check their emails. These answers allowed us to provide the opportunity to frame the subsequent request to sort several emails from the spam folders as a task towards assessment of suspicious emails that are not expected to be phishing or spam by default just because they have been filtered as such by the email providers. Once in the folder, we allowed our participants to select any emails they wanted, but we had to ensure a higher probability that each participant would encounter phishing, spam, or both emails. 

\begin{figure}[!h]
\centering
\includegraphics[width=0.8\linewidth]{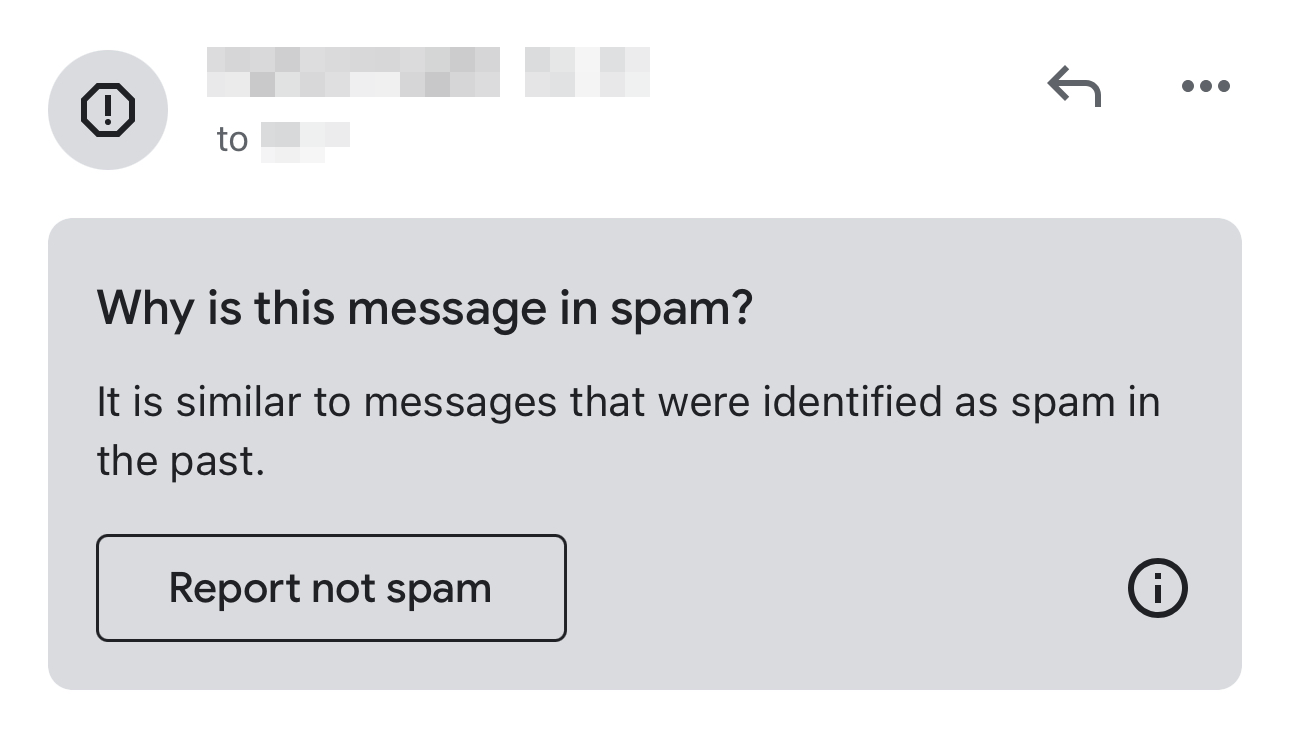}
 \caption{Spam warning banner} 
\label{fig:gmail-spam}   
\end{figure}

\begin{figure}[!h]
\centering
\includegraphics[width=0.8\linewidth]{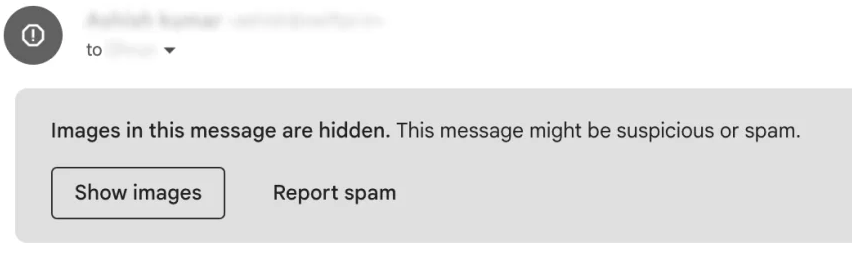}
 \caption{Hidden images warning banner} 
\label{fig:gmail-images}   
\end{figure}

\begin{figure}[!h]
\centering      
\includegraphics[width=0.8\linewidth]{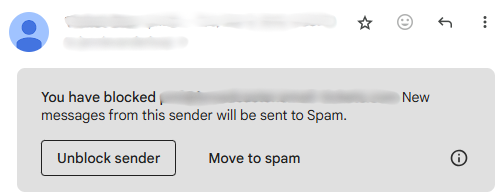}
 \caption{Block sender warning banner} 
\label{fig:blocksender}   
\end{figure}

Google varies the interactive warnings relative to the nature of the suspicious email. Figure~\ref{fig:gmail-spam} shows the banner for \textit{spam} messages, in gray, that instead of signaling the nature of email~\cite{Wogalter2002} it frames the warning through a question that invites the user to go through the suspicious email and ``report it not spam'' if they disagree with the labeling choice by Google (based on the ``similar passages identified in the past''). A variant of this banner, also in gray, is shown in Figure~\ref{fig:gmail-images} where Google has taken action to ``hide the images'' as known elements of deception that conceal suspicious URLs, scripts, or malware, warning the user that ``this message might be suspicious or spam.'' A user here has the option to either ``see the images'' or help the automated detection by confirming the message is actually spam. If a user has blocked a sender of suspicious emails, Google also displays the banner shown in Figure~\ref{fig:blocksender} that tells the user the ``subsequent emails would be sent in spam'' and offers the option to either ``unblock the sender'' or move the existing messages from this sender to their spam folder. 

\begin{figure}[h]
\centering
\includegraphics[width=0.8\linewidth]{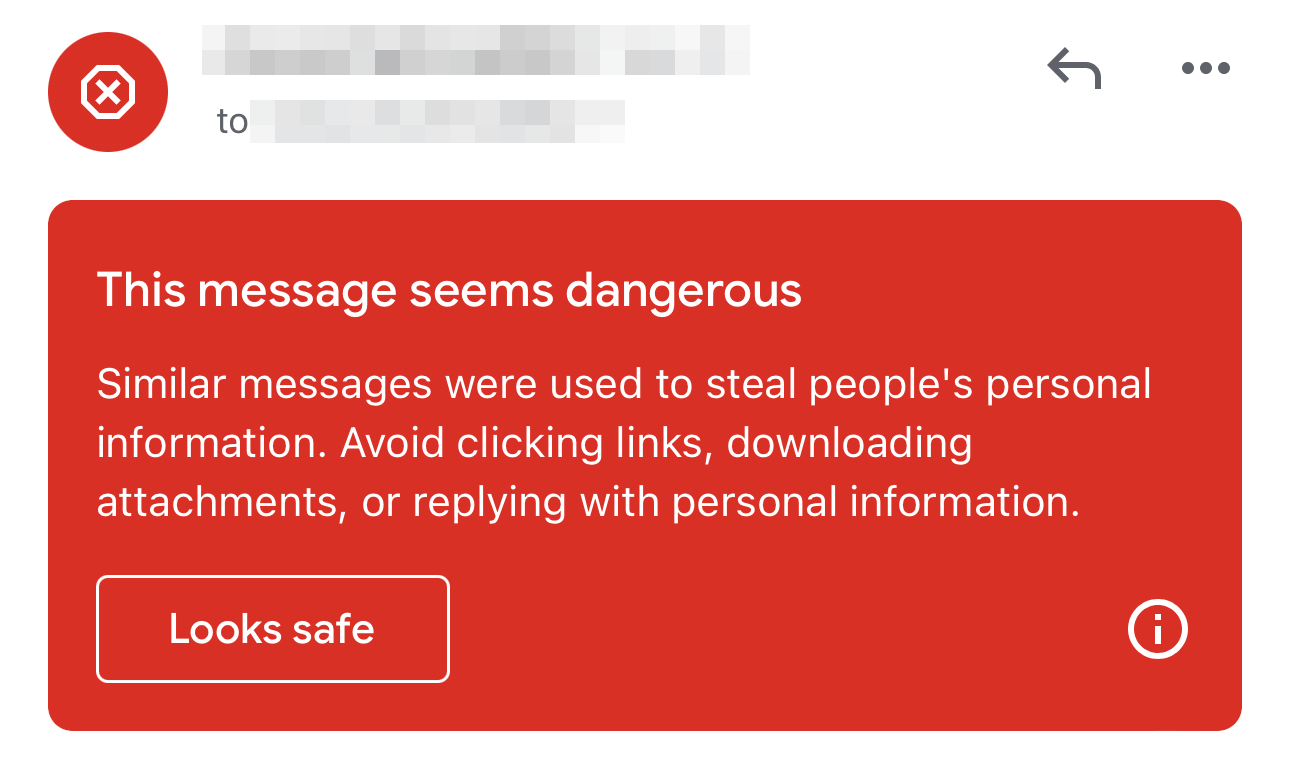}
 \caption{Phishing warning banner} 
\label{fig:gmail-phish}   
\end{figure}

\begin{figure}[h]
\centering
\includegraphics[width=0.9\linewidth]{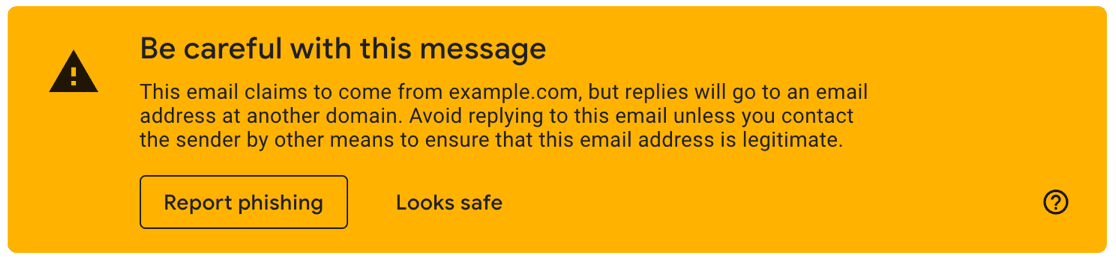}
 \caption{Report phishing warning banner} 
\label{fig:reportphishing}   
\end{figure}

Figure~\ref{fig:gmail-phish} shows the banner for phishing messages, in bright red, that not only alerts the user that the email ``is dangerous'' but also explains the perils of ``clicking on a link or downloading an attachment'' (personal information stolen). If a user is certain that this email is ``safe'', Google offers an option for them to ignore the warning and report it as a false positive. Figure~\ref{fig:reportphishing} shows an alternative banner for \textit{phishing} messages where Google cannot determine with high certainty that the email might be phishing and urges the user to be ``careful with the message.'' If the user, presuming they are versed in spotting phishing cues, determines the email is indeed phishing, then the warning offers the option to report it as such, otherwise the user could proceed and deem the message safe.

While a prior check of multiple researcher-controlled email addresses revealed to us that encountering a spam email with the associated banner would not be a problem, that was not the case for encountering a phishing email in the participants' spam folders. To ensure that this would happen for the purpose of the study, we have prepared a couple of \textit{false positive} suspicious emails -- one with a suspicious URL and one with a suspicious attachment that we knew the filters would assign some of the aforementioned warnings and move them the spam folder. We decided to use false positive suspicious emails, that is, an email which was \textit{not} suspicious but was classified as such by Google anyhow in order to avoid exposing our participants to greater than minimal risk with other, \textit{true positive} suspicious emails.

\begin{figure}[h]
\centering
        \includegraphics[width=0.8\linewidth]{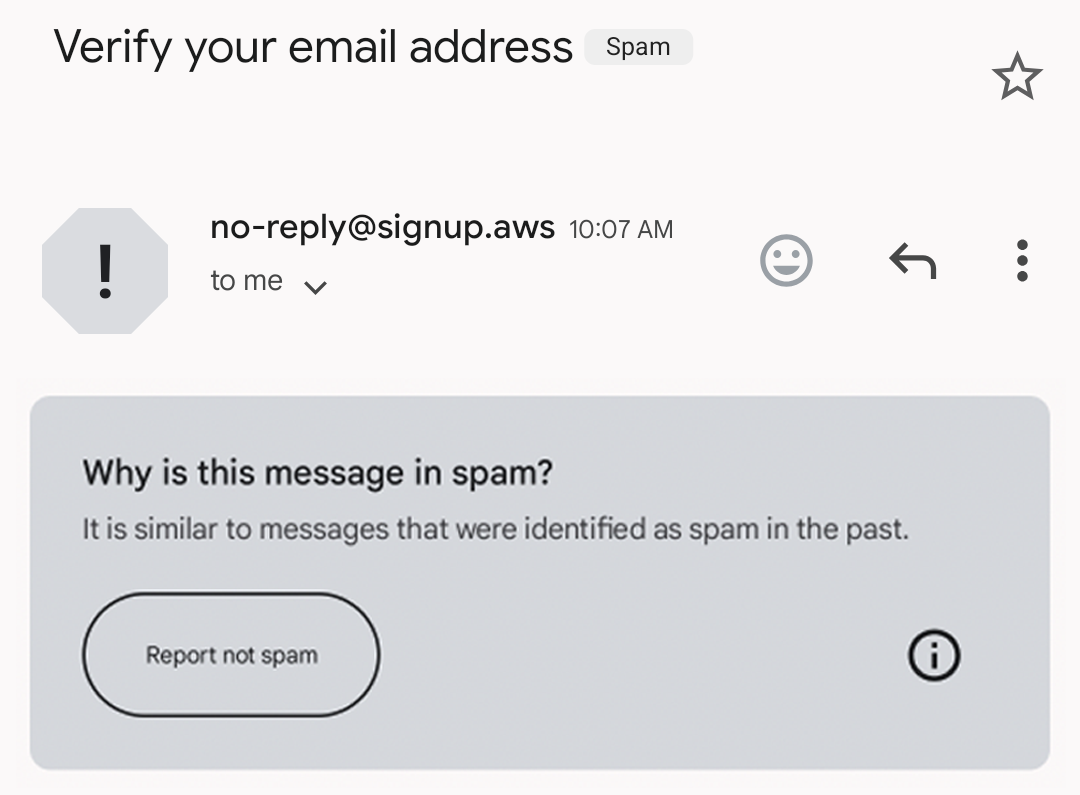}
 \caption{The \textit{false positive} AWS sign-up email} 
\label{fig:link-false-positive}   
\end{figure}

\begin{figure}[h]
\centering
        \includegraphics[width=0.8\linewidth]{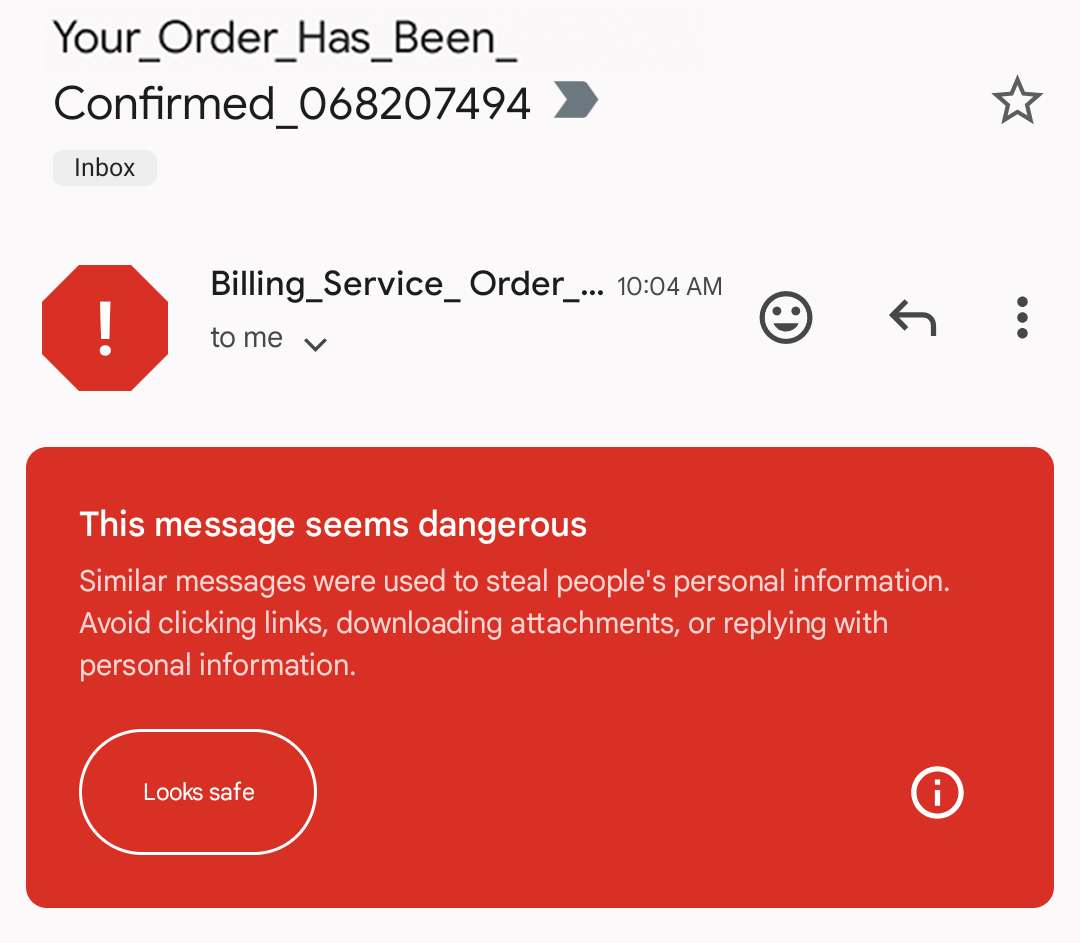}
 \caption{The \textit{false positive} invoice attached email}  
\label{fig:attachment-false-positive}   
\end{figure}

This means the emails could not cause any harm to users as they did not actually contain malicious links (the actual URL included resolved to a legitimate Amazon Web Services, (AWS) page) and the attachment was just a blank Word document. We created the first \textit{false positive} suspicious email by initiating an AWS account creation verification email, shown in Figure~\ref{fig:link-false-positive}. We targeted half of our participants with this (10 in the Amazon Vision Pro group and 10 in the Meta Quest 3 group). The other balanced half we targeted with another \textit{false positive} suspicious email that contained a blank word document that was part of a ``here is your invoice'' pretext that ``confirms a recently placed order'' as shown in~\ref{fig:attachment-false-positive}. Both \textit{false positive} suspicious emails were directed on the day before participation to the participants' emails they used to sign up for the study (we didn't want to tip them off that something might be amiss if we sent the email right before their session).

We chose these emails as we encountered them in our own spam inboxes after we did a personal AWS initiation and searched for emails with attachments. We refer to it as ``mild deception,'' as our participants did not know we were the ones who instigated the sending of this email to their addresses. We assumed the participants might already have spam, but we were less sure about them having suspicious emails containing links and attachments sitting in their spam folders so we could test all the interactive warning banner variants. Both of the \textit{false positive} emails were necessary to invoke a realistic scenario where our participants access either a spam or a phishing banner warning assigned to an email because such an occurrence might not happen frequently enough to be reasonably observed as part of the participants' typical email engagement during the study. 


We were aware that the classification of untrustworthy emails was predicated on the individual's email correspondence and behavior, and we expected that we might encounter a case where both the \textit{false positive} suspicious emails might not end up in participants' spam folders or get assigned any interactive warning. For those cases, we decided to proceed only with what they had in their spam folder as emails addressed to them without going to their main inbox or attempting to ask them to perform additional steps. Our IRB has approved the study, and we used an extensive debriefing (see Appendix \ref{sec:debrief}) in which we pointed out our methodology, discussing any events during the interviews that might affect the participants' future engagement with suspicious emails. We believe that our methodological approach is appropriate because we strike a good balance between the acceptability/manageability of our participants to participate under realistic conditions while we grasp their real-time experiences with their personal email correspondence through a VR headset.

\begin{figure}[!h]
\centering

\includegraphics[width=0.9\linewidth]{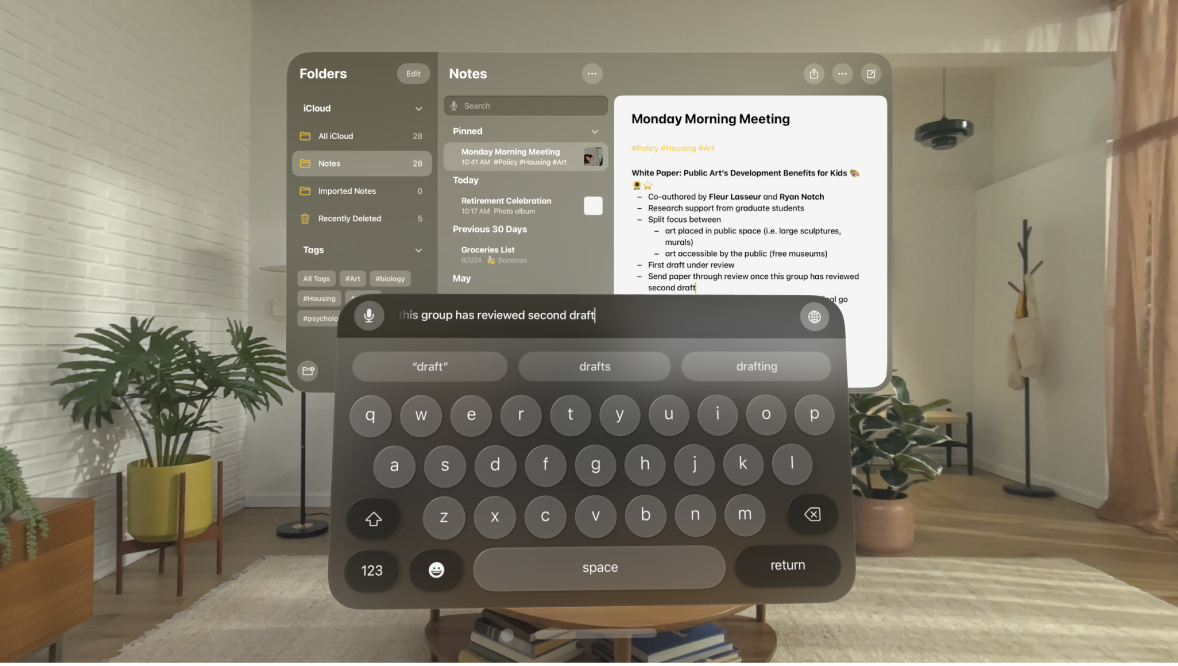}
 \caption{Apple Vision Pro Email Application \cite{Apple2024}} 
\label{fig:applelogin}   
\end{figure}

\begin{figure}[!h]
\centering

\includegraphics[width=0.9\linewidth]{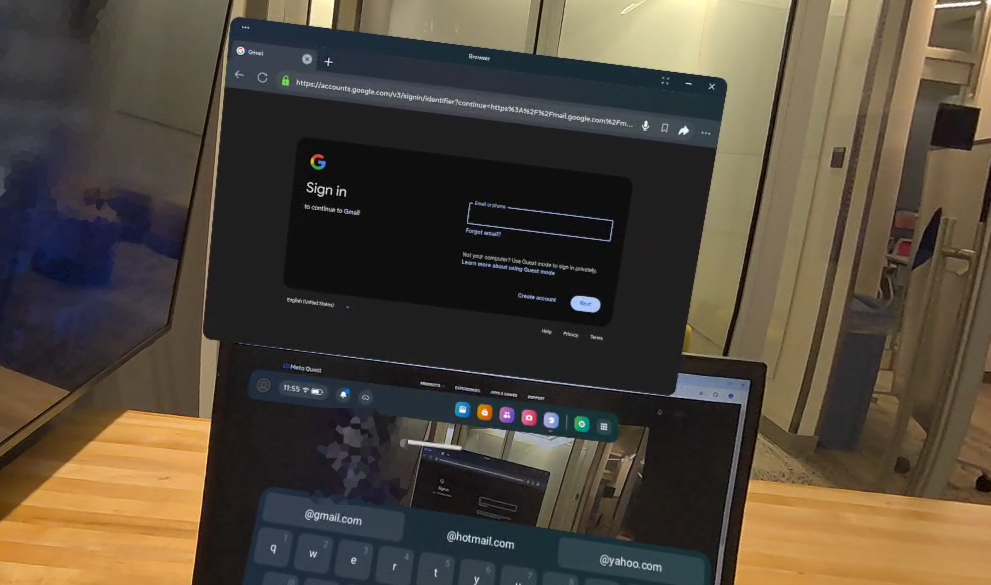}
 \caption{Meta Quest 3 Gmail through Chrome Browser} 
\label{fig:metalogin}   
\end{figure}

Participants were randomly assigned to two groups where they used either the Apple Vision Pro device to check their email via Apple's web application shown in Figure~\ref{fig:applelogin} or the Meta Quest 3 device to check their email via the Google Chrome web browser shown in Figure~\ref{fig:metalogin}. As of the study, Meta has not offered a dedicated email application available to use with the Meta VR, technology so we decided to use the standard browser access to Google email. 

\subsection{Participant Recruitment}
We recruited participants who have used a VR headset and regularly use web/email clients. They had to be individuals 18 years of age or older who had internet access on their own device, client, and browser and were English-speaking and literate. We recruited potential participants through the University's research participant system. We choose to work with students in our University for several reasons. First, students use emails for productivity regularly and are the target population that would mostly adopt VR for such tasks~\cite{Vigderman2024}. Second, across the entire population in the US, college students are the age that mostly uses or has used VR before (gaming, for example)~\cite{Atopia2023, Petrov2024}. And third, we were limited to using the (expensive) VR sets in a physical collaborative space so we could reasonably invite students to such a place on our campus. 

We used a formal email approved by our IRB (see Appendix \ref{sec:recruit}) to approach each of the potential participants. The only stipulation we had was that they use Calendly to sign up for the study and provide their email so we could send them reminders for their study appointment. We arranged in-person interviews with interested respondents on a rolling basis, requesting that each participant has access to any device they require to perform multi-factor authentication verification to access their email. We had no access to their email and we did not record any of their authentication primitives. The interviews were confidential, recorded through an audio-only Zoom session (for the purpose of generating a transcript), lasted on average 30 minutes, and students were provided with extra credit for a class that participates in our University's participant pool program.

\subsection{Trust and Ethical Considerations}
As this was a study in realistic settings and concerning the participants' \textit{own} email addresses, it was important for us to establish trust and assurances about the goals of the study and the safeguard protections we had in place. We first obtained verbal consent both before we started the audio recording and afterward (to have evidence in our transcripts, but also to avoid creating a recording in case a potential participant does not consent, in which we would have thanked them and closed the Zoom session). Next, we communicated that the goal of our study was to capture the ``richness'' of their experiences with sorting their \textit{own} email correspondence in realistic settings. We offered them the option to choose not to participate and to choose which email address they would use, if they had multiple ones. We noted they could select any email they wanted to communicate with us from their spam inbox, and they were free to stop and abandon any question at any point in time if they felt like doing so. Prior to doing any of the tasks and the interview, we told them that they could ask us to stop the interview, stop the recording, or remove any answers or readings at any point in time. 

Only after we received the participants' explicit permission that they were okay with proceeding with the study and accessing the spam folder of their \textit{own} email address, we commenced the audio-only recorded Zoom session and proceeded to complete the tasks and the interviews. We verbally notified each participant when we started each audio recording, and we explicitly told them that they could take as much time as they needed to answer any question. We allowed them to verbalize the process, give comments, complaints, suggestions, and verbalize any other experience related to both the email correspondence and the usability of the VR headset. As the VR headsets might also cause uneasiness or mild dizziness, we also allowed the participants to stop, take off the headset, take a break, and continue when they were ready. We also offered inserts for participants wearing glasses if they needed them for better comfort in using the VR headset, particularly the Apple Vision Pro.  

We pointed out to our participants that they could act on the emails from the study as they ultimately wished (e.g., delete, move to inbox, report, etc.). After we collected their answers, we verbosely debriefed them about the mild deception we used and that we were the ones that initiated a \textit{false positive} suspicious email to their spam inbox (if they received one or chose to verbalize one during the study). We pointed participants to general suspicious email resources if they wished to further raise or check their awareness~\cite{cisa-phishing}. To ensure we obtained a correct understanding of their experiences and recommendations, we reviewed the main points we captured during the interview and clarified any misunderstandings we might have. We also sent a draft of our paper to our participants for feedback.

We employed lengthy explanations to ensure our participants that we were not involved with the filtering nor with the formatting and implementations of the email banner warnings they saw during the study or in the past. We were also careful not to appear in favor nor support of particular types of warnings in order to maintain full researcher impartiality. We communicated that our ultimate goal is to create meaningful detection of phishing when these banner warnings are present in VR headsets. We pointed out that this goal, however, doesn't prevent from misusing our findings or misinterpreting them in making compromises for suspicious email detection, implementation of VR specific email productivity applications, or abandoning the use of VR for productivity altogether.

\subsection{Data Collection}
Initially, the interview transcripts from our Zoom sessions were not anonymized, but we removed any names and references to individual participants and deleted the audio recordings altogether. The transcripts, assigned only with a participant number in the order of participation, were stored on a secure server that only the researchers had access to. Each interview was done with open-ended questions, listed in the interview script (see Appendix \ref{sec:script}). Due to the nature of the study, not all the participants used a similar set of emails, and in some cases, the email providers either entirely filtered out the \textit{false positive} suspicious emails (shown in Figure~\ref{fig:link-false-positive} and~\ref{fig:attachment-false-positive}) or moved them to the main inbox without assigning any banner warnings.

We concluded our recruitment with a sample of 40 participants (20 Apple Vision Pro participants and 20 Meta Quest 3 participants) as we reached thematic/data saturation (i.e., we collected data up to the point where there were fewer surprises in the responses to the research questions and no more emergent patterns). As part of the debriefing process, participants were offered the option to withdraw from the study after finding out about the mild deception (i.e., the \textit{false positive} suspicious email we sent) or no later than 30 days period after the data collection concluded (none of the participants exercised this option). The participant's demographics are given in Table \ref{tab:demographics}.

\begin{table}[tbh]
\renewcommand{\arraystretch}{1.5}  
\caption{Survey Demographic Distribution}
\label{tab:demographics}
\centering
\footnotesize
\begin{tabularx}{0.9\linewidth}{|rYY|}
\hline
& \textbf{\faApple \hspace{0.1em} Vision Pro} & \textbf{\faInfinity \hspace{0.1em} Quest 3} \\\hline
\multicolumn{3}{|c|}{\textbf{Gender Distribution}} \\\hline
\textbf{Male} & 11 & 16 \\
\textbf{Female} & 9 & 4 \\\hline

\multicolumn{3}{|c|}{\textbf{Race/Ethnicity}} \\\hline
\textbf{Asian} & 12 & 14 \\
\textbf{White} & 1 & 5 \\
\textbf{Latinx} & 3 & 1 \\
\textbf{Black} & 3 & 0 \\
\textbf{More than One} & 1 & 0
\\\hline

\multicolumn{3}{|c|}{\textbf{Age}} \\\hline
\textbf{18-19} & 1 & 0 \\
\textbf{20-24} & 12 & 14 \\
\textbf{25-29} & 6 & 2 \\
\textbf{30-34} & 1 & 3 \\
\textbf{35+} & 0 & 1
\\\hline

\end{tabularx}
\end{table}

\subsection{Data Analysis}
Since we had to work with a degree of arbitrary selection of emails in our study, we asked our participants to provide lengthy responses to our questions and asked for further clarifications. With the collected data, we performed an inductive coding approach to identify frequent, dominant, or significant aspects of their answers. As suggested in \cite{clarke2015thematic}, we first familiarized ourselves with the data as we had to manually revise each transcript to remove personally identifiable information. Next, we completed a round of open coding for arbitrarily selected two interviews to capture the participants' decision-making process around the tasks they performed in the study. Then we discussed the individual coding schemes and converged on an agreed codebook (see Appendix \ref{sec:codebook}). The codebook captured four main aspects: (i) \textit{reflection} i.e., codes pertaining to the response upon reflection about the study email; (ii) \textit{email assessment, training, and past experience} i.e., codes related to the past experiences of the participants; (iii) \textit{usability redesign} i.e., codes describing the participants' recommendations for usability redesign of the email banner warnings to detect suspicious emails when accessed through virtual reality headsets. 

We specifically did not include inter-rater-reliability calculations in our process, as our goal was to freely explore diverse topics associated with the natural behavior around immersive email interaction through VR headsets. Overall, our process follows the best practices of usable security and privacy research outlined by Ortloff et. al.~\cite{ortloff2023}. Finally, we discussed and interpreted the identified themes respective to our research questions and wrote down the results. For this, we selected example quotations to represent each of the findings \cite{fereday2006demonstrating}. We utilized verbatim quotations of participants' answers as much as possible, emphasized in ``\textit{italics}'' and with a reference to the participant by order of participation and type of headset used where \textbf{A} denotes the use of the \textbf{Apple Vision Pro} and \textbf{M} denotes the use of the \textbf{Meta Quest 3}. For example, \textbf{P2A} is the second participant to use the Apple Vision Pro headset during our study.

%% file: sections/04.results.tex
\section{Results}

\subsection{RQ1: Email Assessment Through VR Headsets}
\subsubsection{Suspicious Email Warnings Encounters}

More than half of the participants -- 11 Apple Vision Pro and 13 Meta Quest 3 -- encountered the spam banner warning shown in~\ref{fig:gmail-spam} as they sorted through emails in their spam folder. They were able to tell that this banner was about a suspicious email, stating they saw ``\textit{the option to 'report not spam' from Google, as a prevention feature of their email client.}'' (\textbf{P5M}). Interestingly, the \textit{false positive} suspicious emails triggered other warning banners for some of the other participants than the one we exemplified in Figure~\ref{fig:link-false-positive} and~\ref{fig:attachment-false-positive}. 

The ``invoice attachment'' email was flagged as a potentially dangerous phishing email \ref{fig:gmail-phish} for one participant using the Apple Vision Pro and one Using the Meta Quest 3. One Apple Vision Pro participant and one Meta Quest 3 participant encountered the email banner warning that notified users about hidden images in the message, shown in Figure~\ref{fig:gmail-images}. One of the participants that encountered the ``invoice attachment'' email encountered the report phishing banner instead (Figure~\ref{fig:reportphishing}), verbalizing that they have been offered the option by Google to ``\textit{report it as a phishing}'' (\textbf{P19A}) if they felt the email is indeed a phishing one. The full breakdown of the suspicious email warning encounters per VR headset type is shown in Table~\ref{tab:warning}.



\begin{table}[!h]
\renewcommand{\arraystretch}{1.8}
\centering
\small
\aboverulesep=0ex 
   \belowrulesep=0ex 

\caption{Suspicious Email Warnings Encountered}
\label{tab:warning}
\begin{tabularx}{\linewidth}{|p{3.7cm}cY|}
\toprule

\textbf{Warnings} & \textbf{\faApple \hspace{0.1em} Vision Pro} & \textbf{\faInfinity \hspace{0.1em} Quest 3} \\\hline

\parbox[c]{1em}{\includegraphics[scale=0.25]{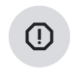}} \hspace{1em} Spam Email & 11 & 13 \\

\parbox[c]{1em}{\includegraphics[scale=0.25]{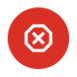}} \hspace{1em} Phishing Email & 1 & 1 \\

\parbox[c]{1em}{\includegraphics[scale=0.39]{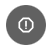}} \hspace{1em} Images Hidden & 1 & 1 \\

\parbox[c]{1em}{\includegraphics[scale=0.40]{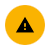}} \hspace{1em} Report Phishing & 1 & 0 \\

\parbox[c]{1em}{\includegraphics[scale=0.55]{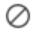}} \hspace{1em} Block Sender & 0 & 1 \\

\hspace{2.3em} No warning & 6 & 4 \\

\hline
\end{tabularx}
\end{table}

Another participant blocked the email address that the \textit{false positive} suspicious email (Figure~\ref{fig:link-false-positive}) came from prior to the interview. When sorting their email, they observed a blocked sender warning as the one shown in Figure~\ref{fig:blocksender}. Referring to the service, \textbf{P15M} stated: ``\textit{I did block them because I needed them (AWS) but, my email didn't get released from their data, I guess; It says because you have blocked that site, so it's in the spam folder, and it gives me the option to unblock the sender.}'' Ten participants did not see an interactive banner warning in the first place for both the \textit{false positive} suspicious emails. Some of the participants were confused by the inconsistency of the warning banners and the fact that not all emails in their spam folders have warnings assigned to them. For example, \textbf{P11A} encountered a marketing email that they ``\textit{wanted it to go to spam; Usually, I get a warning, but now there is none.}'' Even though they found the \textit{false positive} suspicions email with an attachment in their spam folder, \textbf{P13A} commented that``\textit{it doesn't have any warnings, and I'm glad it went here; I wouldn't have really opened it, anyway.}''

\subsubsection{Cues for Accessing Suspicious Emails}
Thirty-five out of all 40 participants (88\%) said they usually relied on elements in the email itself to determine if it was spam or phishing, as shown 
Table~\ref{tab:dealing}. We further broke down these deception cues in Table ~\ref{tab:usability}. The \textbf{context} of the email (the pretext and the circumstances in which it was sent) was the most dominant cue that our participants relied on when sorting suspicious emails. Here, participant \textbf{P7A} stated that emails where ``\textit{the subject line has too many exclamations or the message required an urgent attention to something}'' is a dead giveaway of a phishing email. The second most cited cue was the \textbf{unknown sender}, in addition to other cues. For example, \textbf{P15M} pointed to the ``\textit{the sender's email address, and also the subjects and the attachments}'' as the three elements that indicate an email should be avoided.

\begin{table}[!h]
\renewcommand{\arraystretch}{1.8}
\centering
\small
\aboverulesep=0ex 
   \belowrulesep=0ex 

\caption{Dealing with Suspicious Emails}
\label{tab:dealing}
\begin{tabularx}{\linewidth}{|p{3.0cm}YYYYYY|}
\toprule

\textbf{Assessment Method} & \multicolumn{2}{c}{\textbf{Structure}} & \multicolumn{2}{c}{\textbf{Logic}} & \multicolumn{2}{c|}{\textbf{Grammar}} \\\hline

 & \faApple & \faInfinity & \faApple & \faInfinity & \faApple & \faInfinity \\\hline

\parbox[c]{1em}{\includegraphics[scale=0.25]{figures/spam-icon.png}} \hspace{1em} Spam Email & 10 & 11 & 5 & 6 & 5 & 6 \\

\parbox[c]{1em}{\includegraphics[scale=0.25]{figures/phish-icon.png}} \hspace{1em} Phishing Email & 1 & 1 & 0 & 0 & 0 & 0 \\

\parbox[c]{1em}{\includegraphics[scale=0.39]{figures/images-icon.png}} \hspace{1em} Images Hidden & 1 & 0 & 0 & 1 & 1 & 0 \\

\parbox[c]{1em}{\includegraphics[scale=0.40]{figures/careful-icon.png}} \hspace{1em} Report Phishing & 1 & 0 & 0 & 0 & 0 & 0 \\

\parbox[c]{1em}{\includegraphics[scale=0.55]{figures/block-icon.png}} \hspace{1em} Block Sender & 0 & 1 & 0 & 0 & 0 & 0 \\

\hspace{2.3em} No warning & 6 & 3 & 3 & 3 & 0 & 1 \\

\hline
\end{tabularx}
\end{table}

Interestingly, the third most dominant cue was the \textbf{interactive warnings} themselves. \textbf{P20M} indicated that the banner ``\textit{gets me right into the mindset to think about whether or not the email is spam.}'' Our participants also relied on the presence (or absence) of \textbf{images} in the email body to determine if an email could be spam or phishing. Participant \textbf{P2A} was concerned an email may be dangerous when ``\textit{it appears that the sender did not finish putting in all the images}.'' Ten participants used the \textbf{formatting} of the email body or ``\textit{the way sentences were structured, written, and organized}'' (\textbf{P9A}) to determine that a suspicious email does not follow the typical format that includes a greeting, a main part, and a signature element. Two participants indicated that they assess the \textbf{links} in an email in detail. For example, participant \textbf{P11M} felt that an email containing a link with a domain that seems ``\textit{random, long, and has dispersed special characters}'' is a clear sign of phishing. 


\begin{table*}[!h]
\renewcommand{\arraystretch}{1.8}
\centering
\small
\aboverulesep=0ex 
   \belowrulesep=0ex 

\caption{Suspicious Emails -- Cues}
\label{tab:usability}
\begin{tabularx}{\linewidth}{|p{3.0cm}YYYYYYYYYYYYYY|}
\toprule

\textbf{Cues} & \multicolumn{2}{c}{\textbf{Context}} & \multicolumn{2}{c}{\textbf{Sender}} & \multicolumn{2}{c}{\textbf{Warning}} & \multicolumn{2}{c}{\textbf{Images}} & \multicolumn{2}{c}{\textbf{Formatting}} & \multicolumn{2}{c}{\textbf{Attachments}} & \multicolumn{2}{c|}{\textbf{Links}}\\\hline

& \faApple & \faInfinity & \faApple & \faInfinity & \faApple & \faInfinity & \faApple & \faInfinity & \faApple & \faInfinity & \faApple & \faInfinity & \faApple & \faInfinity \\\hline

\parbox[c]{1em}{\includegraphics[scale=0.25]{figures/spam-icon.png}} \hspace{1em} Spam Email & 9 & 12 & 4 & 4 & 2 & 5 & 6 & 4 & 3 & 5 & 0 & 0 & 1 & 1 \\

\parbox[c]{1em}{\includegraphics[scale=0.25]{figures/phish-icon.png}} \hspace{1em} Phishing Email & 1 & 1 & 1 & 1 & 1 & 1 & 0 & 0 & 0 & 1 & 0 & 1 & 0 & 0 \\

\parbox[c]{1em}{\includegraphics[scale=0.39]{figures/images-icon.png}} \hspace{1em} Images Hidden & 1 & 0 & 1 & 0 & 1 & 1 & 1 & 1 & 1 & 0 & 0 & 0 & 0 & 0 \\

\parbox[c]{1em}{\includegraphics[scale=0.40]{figures/careful-icon.png}} \hspace{1em} Report Phishing & 1 & 0 & 1 & 0 & 1 & 0 & 1 & 0 & 0 & 0 & 0 & 0 & 0 & 0 \\

\parbox[c]{1em}{\includegraphics[scale=0.55]{figures/block-icon.png}} \hspace{1em} Block Sender & 0 & 0 & 0 & 1 & 0 & 1 & 0 & 0 & 0 & 0 & 0 & 0 & 0 & 0 \\

\hspace{2.3em} No warning & 5 & 2 & 3 & 0 & 0 & 0 & 0 & 0 & 0 & 0 & 2 & 0 & 0 & 0 \\

\hline
\end{tabularx}
\end{table*}

Participant \textbf{P13A} indicated phishing attempts ``\textit{happen a lot}'' to them took a more detailed approach, analyzing multiple cues of the email together to decide about its nature:

\begin{quote}
    ``\textit{So I think a lot of it is like the subject of the email, and also the like, the place that it's coming from, like the name of the sender and the email of the sender. Because a lot of the time, what I've noticed with spam emails is that it's just a bunch of letters and numbers in the email. So it's very obvious it's not legit. That's the first thing I go to. And if they have some weird links or like attachments, that's a big sign.}'' 
\end{quote}

While some of the participants who encountered that the spam warning banner relied on it to alert them the email was spam, most felt they needed to do more to make the determination because ``\textit{it says, report not spam, but it doesn't really give you any other information so you don't know what to do with that}'' (\textbf{P8A}). To determine if the email was actually spam, 21 of the 24 participants who encountered the spam banner warning indicated they used the context of the email and ``\textit{had to read through and then search for anything phishy about it}'' (\textbf{P15M}).
Logical cues, such as the improbably of an email request, were an important determination factor for almost half of the participants. 

They identified scenarios with improbable situations, indicating they considered an email phishing if the sender said ``\textit{I'm in danger; I need money; I'll loan you money first; They randomly want to give away things}'' (\textbf{P5M}). Other participants found an email concerning when these two deception components were used together, for example ``\textit{the emails are like really pushy saying you're pre-approved for a credit card or you are automatically entered in this competition}'' (\textbf{P9A}). Participants who encountered no warning used logical cues as often as email elements, identifying an email as suspicious as it was ``\textit{pushing urgency by saying that something happened, even though it didn't}'' (\textbf{P9M}). 

Our participants also found grammatical cues as helpful as logical ones. They encompassed not only errors in grammar but also typos, misspellings, and out-of-order symbols~\cite{Bryant2023}. Those who used grammatical cues scrutinized the details of the text within the email for ``\textit{spelling, font, things like that, special symbols}'' (\textbf{P17A}) and were tipped off ``\textit{if there's typos, that's a big red flag}'' \textbf{P12M}. Others generalized that they ``\textit{don't think the hackers have good grammar}'' (\textbf{P3M}) and saw this as a way to determine if an email sender was attempting to be deceitful.

\subsection{RQ2: Actions Take on Suspicious Emails}
\subsubsection{Avoiding a Phish/Spam} 

During the session, participants were asked about the actions they would take with spam emails. Because the \textit{false positive} suspicious email in either variant was sent prior to their participation session, some participants had already taken action on this email but recalled what action they took. For example, \textbf{P10M} knew they had deleted our email because they ``\textit{make sure that all spam and all my junk is emptied regularly}'' We anticipated that this might occur, given the realistic settings, but nonetheless deemed this a valuable result in a realistic response demonstrating how the participant truly interacts with their spam emails. Table~\ref{tab:actions} shows the actions the participants took based on the warning message they encountered. 

\begin{table*}[!h]
\renewcommand{\arraystretch}{1.8}
\centering
\small
\aboverulesep=0ex 
   \belowrulesep=0ex 

\caption{Actions Taken During Email Sorting}
\label{tab:actions}
\begin{tabularx}{\linewidth}{|p{3.0cm}YYYYYYYYYYYY|}
\toprule

\textbf{Actions} & \multicolumn{2}{c}{\textbf{Ignore}} & \multicolumn{2}{c}{\textbf{Investigate}} & \multicolumn{2}{c}{\textbf{Unsubscribe}} & \multicolumn{2}{c}{\textbf{Delete}} & \multicolumn{2}{c}{\textbf{Report}} & \multicolumn{2}{c|}{\textbf{Block}}\\\hline

& \faApple & \faInfinity & \faApple & \faInfinity & \faApple & \faInfinity & \faApple & \faInfinity & \faApple & \faInfinity & \faApple & \faInfinity \\\hline

\parbox[c]{1em}{\includegraphics[scale=0.25]{figures/spam-icon.png}} \hspace{1em} Spam Email & 8 & 8 & 2 & 2 & 1 & 2 & 0 & 1 & 0 & 0 & 0 & 0 \\

\parbox[c]{1em}{\includegraphics[scale=0.25]{figures/phish-icon.png}} \hspace{1em} Phishing Email & 1 & 1 & 0 & 0 & 0 & 0 & 0 & 0 & 0 & 0 & 0 & 0 \\

\parbox[c]{1em}{\includegraphics[scale=0.39]{figures/images-icon.png}} \hspace{1em} Images Hidden & 1 & 1 & 0 & 0 & 0 & 0 & 0 & 0 & 0 & 0 & 0 & 0 \\

\parbox[c]{1em}{\includegraphics[scale=0.40]{figures/careful-icon.png}} \hspace{1em} Report Phishing & 1 & 0 & 0 & 0 & 0 & 0 & 0 & 0 & 0 & 0 & 0 & 0 \\

\parbox[c]{1em}{\includegraphics[scale=0.55]{figures/block-icon.png}} \hspace{1em} Block Sender & 0 & 0 & 0 & 0 & 0 & 0 & 0 & 0 & 0 & 0 & 0 & 1 \\

\hspace{2.3em} No warning & 1 & 0 & 2 & 1 & 1 & 1 & 0 & 2 & 1 & 0 & 1 & 0\\

\hline
\end{tabularx}
\end{table*}

Upon reviewing the \textit{false positive} email, 17 participants whi encountered the spam banner indicated that they were satisfied with the email going to their spam folder and said they ``\textit{would not revert this email back to their inbox}'' (\textbf{P2M}). Four of the participants that encountered the spam warning conducted a cursory investigation, saying they would ``\textit{open up an email, but I won't really dive too deep into it if there's something sketchy or off with it}'' (\textbf{P14M}). Two participants indicated they would go to the end of the email and``\textit{typically just unsubscribe.}'' (\textbf{P1M}), and one said they would ``\textit{scroll for 10 seconds and, yep, delete them all}'' (\textbf{P7M}) as Google offers the possibility to instantly remove all the spam. 

The two participants who encountered the phishing email warning took additional precautions when evaluating the emails and elected not to interact with them. One of these participants, \textbf{P11M}, stated that when they see that banner, ``\textit{usually [they] don't click on it; [they] just ignore the email altogether}.'' The two participants who encountered the ``hidden images'' warning indicated that they wouldn't interact with the email, and they found it convenient. One of them, \textbf{P17A} stated: ``\textit{I like that the banner is makes you have to opt in to view images; No skin off my back to click the little button if I think it's a legitimate email.}'' The participant that encountered the report phishing banner warning indicated that they ``\textit{would say it makes you think twice about logging or clicking on any links}'' (\textbf{P19A}) and opted not to interact with the email as well. 

We observed that, in both VR groups, the participants who did not encounter any warning during the sorting task were more likely to closely inspect and interact with the elements of the email. When dealing with potential spam emails, three participants felt they needed to ``\textit{check to see how legit it is}'' (\textbf{P9M}). Two participants in this group said they would ``\textit{unsubscribe it from my inbox.}'' (\textbf{P18A}), and two said they would delete the email. Only two participants in the group said that when they came across a spam email they, ``\textit{would just ignore it}'' (\textbf{P14A}).

\subsubsection{Falling for a Phish}
Three of the participants that encountered our test \textit{false positive} suspicious emails erroneously clicked on the link or opened the attachment. Two Apple Vision Pro participants clicked on the link in the email, which brought them to a legitimate Amazon Web Services (AWS) page (as the email was legitimate in nature). They acknowledged that they did so and stated shortly that they did it ``\textit{unintentionally}'' (\textbf{P15A}), but both of them were hesitant to go into more detail. We did not probe them further as both of them felt uncomfortable that they fell for the phish (we observed their body language, avoiding eye contact, and looking to proceed with the other questions in the interview). One Meta Quest 3 participant erroneously opened our blank Word document attached to the email, vocalizing what appeared to be an accidental misclick as ``\textit{Oh, shit! I actually opened the document!}'' (\textbf{P2M}).

\subsection{RQ3: Usability Recommendations}

\subsubsection{Warning Banner Usability Recommendations}
While most participants heeded the warning in the banner, many said they found the warning messages to be only partially usable. Half of the participants indicated that they had to go into their spam folder to look for important emails, including one-time passwords, newsletters they opted in for, and contacts from potential employers, resulting in a need to assess the legitimacy of the emails in their spam folder. They thought it was helpful to get some kind of warning but that the warning banner could be improved to help users better understand what elements of the email to look at and what actions to take with the email. Table \ref{tab:bannerrecommend} shows the usability recommendations made by participants after viewing the warning message they encountered.

Thirty participants, or 75\% that encountered the standard spam warning, were underwhelmed by the message and color coding they encountered. Four of the participants in this group had recommendations of ways to improve the banner to alert them of potential dangers. For example, \textbf{P8M} reflected that:

\begin{quote}
``\textit{The top message should say, `this is \textit{definitely} spam.' So then I might be a little more keen with what I'm looking at and what I am doing. I didn't feel that I have to be careful. It just says this could be a spam. Then, okay, let me go through and when I saw the scary things at the bottom, I was, 'Oh, my God'.}'' 
\end{quote}


\begin{table*}[!h]
\renewcommand{\arraystretch}{1.8}
\centering
\small
\aboverulesep=0ex 
   \belowrulesep=0ex 

\caption{Warning Banner Usability Recommendations}
\label{tab:bannerrecommend}
\begin{tabularx}{\linewidth}{|p{3.0cm}YYYYYYYYYY|}
\toprule

\textbf{Recommendations} & \multicolumn{2}{c}{\textbf{Color Coding}} & \multicolumn{2}{c}{\textbf{Algorithm Control}} & \multicolumn{2}{c}{\textbf{Message Content}} & \multicolumn{2}{c}{\textbf{Pop-up}} & \multicolumn{2}{c|}{\textbf{Risk Indicator}} \\\hline

& \faApple & \faInfinity & \faApple & \faInfinity & \faApple & \faInfinity & \faApple & \faInfinity & \faApple & \faInfinity \\\hline

\parbox[c]{1em}{\includegraphics[scale=0.25]{figures/spam-icon.png}} \hspace{1em} Spam Email & 4 & 2 & 2 & 2 & 1 & 5 & 0 & 1 & 0 & 1 \\

\parbox[c]{1em}{\includegraphics[scale=0.25]{figures/phish-icon.png}} \hspace{1em} Phishing Email & 1 & 0 & 0 & 0 & 1 & 0 & 0 & 1 & 1 & 0 \\

\parbox[c]{1em}{\includegraphics[scale=0.39]{figures/images-icon.png}} \hspace{1em} Images Hidden & 1 & 0 & 0 & 0 & 0 & 0 & 1 & 0 & 1 & 0 \\

\parbox[c]{1em}{\includegraphics[scale=0.40]{figures/careful-icon.png}} \hspace{1em} Report Phishing & 1 & 0 & 0 & 0 & 0 & 0 & 0 & 0 & 0 & 0 \\

\parbox[c]{1em}{\includegraphics[scale=0.55]{figures/block-icon.png}} \hspace{1em} Block Sender & 0 & 0 & 0 & 0 & 0 & 0 & 0 & 0 & 0 & 1 \\

\hspace{2.3em} No warning & 0 & 1 & 4 & 1 & 1 & 0 & 0 & 0 & 0 & 0 \\

\hline
\end{tabularx}
\end{table*}

Six participants in the group that encountered said that the color of the banner could be updated to fit the immersiveness of the VR, and one thought that colors could be used as a risk indicator. Participant \textbf{P1A} reasoned that when doing email sorting through a VR headset ``\textit{color coding would help a lot, because although [Google] does filter such messages and puts them in another mailbox, overall, it just all looks the same}.'' Participant \textbf{P16M} echoed the sentiments suggesting ``\textit{that the color coding would definitely help because it can be tough to see the difference in there, especially during multitasking}'' (we offered the participants to act naturally and do other tasks in the VR while sorting their emails so we mimic their natural task-switching behavior). 

Here, participant \textbf{P17A} felt that in a VR environment, it would be fitting and useful to have a full palette of warnings (per~\cite{Wogalter2002}). They suggested that Google should implement a ``\textit{green indicator about legitimate emails that's not necessarily a banner but something small that people won't ignore over time}.'' Participant \textbf{P13A} worried about vulnerable populations, saying that color coding would help``\textit{make the suspiciousness more obvious, maybe because I know a lot of people, especially my dad, who's old, so he would definitely press something if he has to open an email through a VR headset}'' (\textbf{P13A}).

The overall satisfaction with Google's sorting and assignment of banners was low. Twenty-nine, or 73\% of participants said they regularly had spam go to their inbox or important emails go to their spam, and most felt that the banner warnings were assigned inconsistently. Four participants in the group that encountered the spam warning variants, and five in the group that encountered no warning wanted more control over the emails and the way Google handles them, especially when one has to access them through a VR headset. Participant \textbf{P20M} suggested that anyone who would want to use their email through a VR headset should go through an ``\textit{explicit training period so they train the engine to respond according to their interactions.}'' As the immersiveness is the key feature that would accelerate the adoption of VR headsets, participant \textbf{P18M} reasoned that such training could be good in the context of ``\textit{allowing people to maneuver email messages with gestures like they would do with real mail}'' (alluding to the action of tearing spam mail or physically discarding it).

There were three participants that thought a ``\textit{three-dimensional pop-up in the VR headset would help}'' (\textbf{P7M}) felt it would be helpful in the addition of the ability to drag, scale, position, and orient various windows simultaneously with both hands. Participants were unsure if it was the responsibility of their email provider, the VR headset technology, or a shared responsibility. Here, participant \textbf{P11M} surmised that ``\textit{if something is labeled as spam in my inbox I can see there would be ways, scaling or immediate decluttering for the headset to make it harder to click on something in the spam or bring your focus that you're about to click on something}''.

\subsubsection{Virtual Reality Usability Recommendations}

While performing the email sorting tasks, 25\% of participants (five Apple Vision Pro and five Meta Quest 3) worried about the hypersensitivity and precision involved in clicking using a VR headset, accurately predicting the accidental ``misclicks'' that actually occurred with three of our participants. As shown in Table~\ref{tab:vrrecommend}, there were differences between the Apple Vision Pro and the Meta Quest 3 recommendations based on the type of device used. The Apple Vision Pro uses vision tracking and, to click on something using this VR headset, the user looks at the object they would like to select and then taps their index finger and thumb together to select it. The Meta Quest 3, on the other hand, has two joysticks with buttons. The user holds the joysticks in their hands, points the cursor at the object, and then clicks a button on the joystick to make a selection.

The Apple Vision Pro participants stated that ``\textit{the touch is a little too sensitive}'' (\textbf{P8A}), that perhaps even after calibration, it would create situations where, in the words of participant \textbf{P1A} it's ``\textit{hard to zero-in on a single button and I would probably accidentally move my head and click on another thing I don't mean to.}'' The Meta Quest 3 participants commented on the noticeable glitches with the precision fix of the VR headset. Participant \textbf{P9M} confessed that ``\textit{while [they] were trying to navigate through all this, [they] almost clicked three times on random other buttons out there}.'' Speaking for the overall experience with the Meta Quest 3, participant \textbf{P11M} explained:

\begin{quote}
``\textit{The thing with this headset is that you have to move the cursor around with your hand, and sometimes that's not as accurate as using a mouse or your finger, so I can see that you could actually then maybe click a bad link from a spam message accidentally. I suppose there must be ways to make the friction, feedback, and response configurable so it feels more natural}''

\end{quote}

\begin{table}[!h]
\renewcommand{\arraystretch}{1.8}
\centering
\small
\aboverulesep=0ex 
   \belowrulesep=0ex 

\caption{Virtual Reality Usability Recommendations}
\label{tab:vrrecommend}
\begin{tabularx}{\linewidth}{|p{1.8cm}YYYY|}
\toprule

\textbf{ } & \textbf{Sensitive Clicking} & \textbf{Precise Clicking} & \textbf{Multi-tasking} & \textbf{Email App} \\\hline

\faApple \hspace{0.1em} Vision Pro & 5 & 0 & 1 & 0 \\

\faInfinity \hspace{0.1em} Quest 3 & 0 & 4 & 1 & 3 \\

\hline
\end{tabularx}
\end{table}

Two participants from each group said that the way the screen is set up is too distracting for them to be able to trust themselves when determining phishing emails while performing other tasks. For example, \textbf{P1M} said that ``\textit{With a VR headset you have more going on, so it's easier not read a warning of thoroughly, as your focus might be on another task or orientation}.'' In response to the multitasking, participant \textbf{P1M} recommended that... 

\begin{quote}
``\textit{When you open your browser or email client, whether the headset company implements this or Gmail, maybe, everything else should be dimmed down by default, so it creates kind of a place where it the focal point of your attention should be just the email. When I was going through my emails, everything else remains in full focus and brightness, So I was easily distracted, you know, I could have very much clicked on something.}''    
\end{quote}

The Apple Vision Pro also has a dedicated email application available in the Apple App Store that participants were able to use to access their email, as shown in Figure~\ref{fig:applelogin}. The Meta Quest 3 does not have a dedicated email application available at the time of our study, and our participants accessed their email provider's website through the Google Chrome web browser, as shown in Figure~\ref{fig:metalogin}. We didn't observe any specific differences between the app and the browser per se (which is a variable for future exploration), but we probed the Meta Quest 3 about the experience of sorting the correspondence through a browser. Three of them felt that their challenges with navigation and the potential to misclick might be resolved with ``\textit{an application which is specifically designed for this kind of space, both for work and especially for gaming}'' (\textbf{P8M}).

%% file: sections/05.discussion.tex
\section{Discussion}
\subsection{Implications: Virtual Reality}
Our results suggest that VR headsets are not immune to the threat of phishing/spamming and could even catalyze such attacks with the very nature of the immersive interaction they offer. In situations where an elaborate observation-based attacks are infeasible~\cite{Arafat2021, Gopal2023, Khalili2024, luo2024eavesdropping, Slocum2023} -- which is expected to be the norm in a working environment or at home where people might use the VR headset for productivity, an adversary could always resort to sending a phishing/spam email and hope to get the full credentials of a target. Apple Vision Pro's eye tracker approach was shown to be exploitable through eye-related biometrics' inferences~\cite{wang2024gazeploit}, but such an attack is headset-specific. A phishing/spamming adversary has no such constraints and could target a user that might switch between VR-headsets easily. This cross-device versatility works in equal capacity to overcome a potential malicious VR trap-in~\cite{yang2024inception}.

\subsection{Implications: Suspicious Email Warnings}
For long, the interplay between phishers/spammers and first-line email detection is defined by the ability of the adversaries to craft emails with clever formatting, spelling, keyword selection, and sender spoofing~\cite{Verizon.2023}. We expect these tactics to remain, but the immersiveness itself offers an opportunity for expanding the space for experimenting with deceptive affordances further. For example, adversaries could include images, fonts, and animation in emails that might distract a user away from the warning, the link, and the overall email. The concept of \textit{external} distractions is already available in medical VR applications aiming to reduce pain and anxiety of patients, for example~\cite{glennon2018use}, and we believe that it is not entirely unlikely that an adversary would resort to \textit{internal} distractions as a mean of deception. Distraction is such an important susceptibility factor that advanced suspicious email warnings incorporate a temporal component or a delay of a few seconds into rendering the email content in order to eliminate possible in-situ distractions~\cite{Mossano2023, Petelka2025}. In a multitasking environment such as VR, such delays might not be possible to prevent phishing/spamming distractions, or worse, they could be of an adversarial advantage.

\subsection{Implications: Suspicious Email Training}
The VR might itself create circumstances where the immersiveness is used for deceptive purposes, but the same property could be equally utilized towards a more noble goal, that is, phishing/spam awareness training and education. The early evidence shows that VR headsets could be effectively used to practice phishing attacks and train users to spot the most prevalent cues of deception~\cite{JansenFischbach2020, Bakker2024}. The interactivity of the VR interface might be an ideal place where novel approaches of training could naturally blend warnings, trainings, and especially gamification for a lasting learning experience against various types of social engineering~\cite{Berens2024}. Furthermore, the personalization property of the VR headsets could equally enable personalized suspicious email training based on users' phishing knowledge and current detection skills~\cite{schoni2024you}. As VR headsets target productivity, we see them as an ideal tool for enterprise level phishing training and simulated exercises that could alleviate the chronic problem of minimal interaction~\cite{Ho2025}.

\subsection{Limitations}

The realistic settings impose several limitations pertaining to our study. A limitation comes from the sample size that prevents the generalization of the results to the entire population of VR device users and email providers other than Gmail. Another limitation is that we sampled English-speaking email users from the United States and used the email banner warnings in Gmail in their English variant. Other language implementations of the warnings and the email client interface, as individuals from countries other than the US, might yield different results than ours. Aside from using only Google Mail (or Gmail), a limitation comes from the fact that we used the implementation version of the respective banner warnings at the time of the study, that is, the last part of 2024. Future changes, adaptations, and improvements in the way Gmail makes its banner warnings usable -- especially for access through VR headsets -- might render our results obsolete (which we sincerely hope will be the case as the VR gets traction for productivity purposes). 

Our participants were university students who were able to participate in in-person sessions. For the time being, this is an expected limitation imposed on any empirical VR research due to the need for managing expensive VR headsets, which hinders the ability to recruit a wide variety of participants through crowdsourcing methods and remote studies. While remote studies through social VR platforms have been conducted in the past~\cite{Saffo2021}, a generalization of this methodological approach is currently limited, especially for mild deception and observational studies like ours that demand close proximity of the researchers for the purpose of safeguarding and prevention of harm. The time and space of participation was also a limitation as we used a shared collaborative space. An actual work or work-from-home physical environment might affect the degree of VR immersiveness and, as such, affect how a user approaches a suspicious email. 

The current version of the VR headsets our participants used as part of the study poses yet another limitation, as any new features might transform how VR device users access and use the email banner warnings and, with that, affect the overall findings. Additionally, we only used two brands of VR headsets. The Apple Vision Pro and the Meta Quest 3 were selected for their novelty and market position at the time of our study. The Apple Vision Pro was released in the early part of 2024 and was highly anticipated. The Meta Quest 3 was released in late 2023, and as of the time of our study, holds the largest market share for VR devices of more than 80\% in the US~\cite{Zreality2024}. The Apple Vision Pro and the Meta Quest 3 were considered two of the best VR headsets available during our study, being selected as the device with the best AR/VR interface and the best standalone VR headset, respectively~\cite{Greenwald2024}.

A structural limitation is the choice of using the participants' spam inbox instead of their own inbox. This research setup might have primed our participants with a hint that these emails were not to be trusted, even though we took measures to minimize such a hint before participants did the tasks in our study. The priming, however, was a necessary compromise to introduce and test a novel methodology that brings experiments close to the real interaction with suspicious emails in VR environments and away from the laboratory settings typical for empirical studies with suspicious emails. We cannot fully generalize our results to individuals' main inboxes because we were not permitted by our IRB to phish, spam, or tamper with Gmail's filtering rules due to greater than the minimal risk to them. Another related limitation comes from the choice of the \textit{false positive} emails we used. The particular sender and the email request itself, as well as the nature of the attachment, might have contributed to a phishing tip-off in a greater capacity/effect than the warning itself. Other phishing emails might generated other warnings and have yielded different assessment/action results within the VR environment. 

Similarly, a limitation comes from the choice of spam emails our participants arbitrarily selected in our study, as any other message or an encounter with a legitimate email marked as spam might have caused a different behavior around the banner warnings. Though we left our participants sufficient time and support to engage with the emails and the email banner warnings through the technology of their choice, this might not have been sufficient for them to formulate a more informed expression about their overall suspiciousness assessment and ultimate decision about it. Despite all these limitations, and similar to qualitative studies in general, our study nonetheless provides rich accounts of individuals' lived experiences with suspicious emails using VR devices, which studies in laboratory settings hardly offer.

\subsection{Future Work}
Due to the sampling limitations, we were restricted to tech-familiar, relatively young university students. While the realm of VR applications is expanding, particularly towards productivity, VR itself is seen as transformative from an accessibility perspective that could revolutionize the way people with disabilities interact with digital content~\cite{Mott2019}. Suspicious emails and the associated warnings are inaccessible, for example, for blind or low vision users~\cite{Yu2023, Sharevski2024-usenix}. In our future work, we will focus on developing and testing the possibility of adding textual metadata and haptic feedback as part of the interactive warnings in order to allow for people with vision disabilities the opportunity to use VR for productivity in a similar capacity as their sighted counterparts. Equally, we are interested in exploring the possibility of a social VR application where blind and low vision people could use the aid or sighted counterparts in sharing experiences and advice about suspicious emails. We also plan to address the expansion of the VR towards the older adult population as one particularly vulnerable to social engineering, given the positive experiences of these users with exposure to virtual reality~\cite{Abeele2021}.


%% file: sections/06.conclusion.tex
\section{Conclusion}
During our study, we had 40 participants (20 Apple Vision Pro users and 20 Meta Quest 3 users) sort their own emails using a VR headset. The sorting was done in the presence of a test \textit{false positive} suspicious email that we covertly sent to the participants prior to the study (and later debriefed them about it). Our goal was to learn how an email user in a realistic VR setting would access a suspicious email, that is, what cues -- such as the context, formatting, URLs, and presence of attachments, along with the Gmail warning banners -- would they use to determine how to act upon it. While our participants utilized the warning banners, they felt they needed improvements to be fully transferable for an immersed user to heed them effectively. Participants also experienced unique VR-related interaction challenges that led three participants to fall for our test suspicious email by clicking a link or opening an attachment. Our findings clearly point to the need for a design of dedicated VR-based suspicious email warnings within a designated email client that would limit the ability of a phishing/spamming adversary to capitalize the immersiveness to their advantage.

%% file: sections/appendix.tex
\clearpage
\appendices

\section{Recruitment Email} \label{sec:recruit}

\noindent \textbf{From}: Researcher's Email \\
\textbf{Subject}: Research Study Participation \\
\textbf{Date}: \\
\textbf{To}: Potential Participant

\vspace{0.5em}

\noindent Hello,

\vspace{0.5em}

\noindent My name is \censor{Filipo Sharevski, Ph.D.}, a professor at \censor{DePaul University's School of Computing and Digital Media}. I am conducting a research study about the experiences with email sorting through Virtual Reality (VR) headsets. \\

\noindent The purpose of the research is to learn more about how people utilizing virtual reality headsets to interact with emails general. I am recruiting only volunteers for this research study and you can participate on your own volition. You are eligible to participate if you are 18 years or above old, you are from United States, you are able to understand and converse in English language, are able to use a VR headset, and you are actively using emails as a mode of communication. \\

\noindent I will ask you several open-ended questions and collect some personal information about you such as age, gender, ethnicity/race, education, computer, and VR headset proficiency. If there is a question you do not want to answer, you may ask us to skip it. Your information will be kept confidential and stored in a secured computer under password protection and with encrypted files. The data will be kept de-identified. The participation will take about 30 minutes. You will receive an extra credit for a successful participation in a class of your choice (selected in the \censor{SONA} system during the sign-up).\\


\noindent You must have an email client and internet connection on your phone. You must be able to understand and converse in English. \\

\noindent If you are interested in participating, please respond directly to this email expressing your interested in the study and your familiarity with virtual reality headsets. My email is \censor{fsharevs@depaul.edu} and my cell phone number is \censor{+1 765-714-9574}. Once I have ascertain your eligibility, we will agree for a timeslot for your IN-PERSON participation in the \censor{DePaul's Loop Campus}\\

%


\noindent I thank you for your time in reading this email. 




\section{Interview Script} \label{sec:script}

\subsection*{Announcement}
This interview is being audio-recorded for research purposes. You may stop the recording at any time. Do you consent to being audio-recorded? Recording starts now.

\subsection*{Questions and Tasks}
\begin{enumerate}
\itemsep 0.5em


\item Which statement best describes your level of experience with virtual reality headsets?
\begin{itemize}
    \itemsep 0.5em
    \item I have tried one out, but never operated it alone.
    \item I have used a headset for gaming.
    \item I have used a headset for functions outside of gaming (ask for types of tasks performed).
\end{itemize}

\item What device do you prefer to read your emails on?
\begin{itemize}
\item Phone (ask for make and OS)
\item Desktop/Laptop (ask for make and OS)
\item Tablet (ask for make and OS)
\item VR Headset (ask for make and OS)
\end{itemize}


\item Can you please check your spam or junk folder. Can you please go over the most recent few emails -- one by one -- you have received in this folder.  

\item Carefully open each of these emails and just review the contents. We would like to ask some questions about this particular experience: 

    \begin{enumerate}
        \item [1.1] Have you noticed anything unusual about these emails? Please specify in as many details as you can. 

        \item [1.2] Have you noticed any warnings, notifications, or labels about these emails? Please specify in as many details as you can. 

        \item [1.3] How do you usually review and decide what to do with these emails?

        \item [1.4] How do these warnings, notifications, or labels affect your opinion about the safety of the email they were substantiated to (e.g. email phishing or not, spam or not, scam or not)?

        \item [] \textit{Note: A baseline of ``suspicious'' and ``safe'' (legitimate) is established, based on the the participant's interpretation of what they think is suspicious (or safe). In case this interpretation differs from the definition of phishing/spam suspiciousness/legitimacy~\cite{cisa-phishing}, a brief explanation of ``suspicious'' and/or ``safe'' (legitimate) is provided. This is done as such to minimize the probability of priming participants for the subsequent tasks.} \label{app:baseline}

        \item [1.5] What cues do you usually use to assess the legitimacy of emails?

        \item [1.6] How often do you see suspicious emails in your inbox?

        \item [1.7] How often do you see legitimate emails in your spam folder?
        
    \end{enumerate}

\item What is your opinion on how your email provider handles suspicious emails? 

\item Have you received any phishing, spam, scam or dangerous email training? 

\item Have you ever been a victim of a successful phishing, spam, or a email scam campaign? If you are comfortable with, please share your experiences with this event(s). [What lessons you have learn from here and how this episode affected your way of dealing with emails afterwards]

\item Have you seen any other types of phishing, spam, scam campaigns delivered over other types of communication than email (e.g. SMS, social media, Discord)?

\item What would you recommend about how these email warnings, notifications, labels should be made to work for individuals utilizing virtual reality headsets?

\item Anything else you want to add on this topic or your experience with warnings about emails?

\item Demographic Questions

\begin{enumerate}
\item How old are you?
\item What race/ethnicity do you identify as?
\item What is your gender?
\end{enumerate}

\end{enumerate}

\section{Codebook} \label{sec:codebook}


\subsection{Reflection}

\begin{enumerate}
\itemsep 0.5em 

        \item \textbf{Banner Message Received} Codes pertaining to the warnings and indicators the participant observed on their spam emails.
        
        \begin{itemize}
            \item \textbf{Why is This in Spam} The participant expressed their spam emails contained a warning indicating that the email was spam and they had to opportunity to report that the email was not spam
            \item \textbf{Images are Hidden} The participant expressed their email contained a warning that the images are hidden for their safety
            \item \textbf{Dangerous Email} The participants expressed that their spam email contained a warning that the email or attachment was potentially dangerous
            \item \textbf{Unblock Sender} The participants expressed that their spam email contained a warning that the sender had previously been blocked
            \item \textbf{No Warning} The participant expressed that their emails did not have any warning message
        \end{itemize}

        \item \textbf{Cues} Codes pertaining to the cues noticed by the participant during their experience with the emails in their spam folder.
        
        \begin{itemize}
            \item \textbf{Warning} The participant expressed their spam emails contained a warning indicating that the email was spam or phishing
            \item \textbf{Unknown Sender} The participant expressed that the email was from an unknown sender
            \item \textbf{Context} The participant expressed that the email was unsolicited or inapplicable
            \item \textbf{Attachment} The participant expressed that there was an attachment that did not fit with the content of the email
            \item \textbf{Formatting} The participant expressed that the email contained characters or grammar inconsistent with their expectations for an email in English
            \item \textbf{Images} The participant expressed that the images did not load properly or were missing
        \end{itemize}

        \item \textbf{Action} Codes pertaining to the actions taken by the participant during their experience with the emails in their spam folder.
        
        \begin{itemize}
            \item \textbf{Leave} The participant indicated they would leave the email in their spam folder and ignore it
            \item \textbf{Investigate} The participant indicated that they would further investigate the elements of the email to determine if it is legitimate
            \item \textbf{Delete} The participant indicated they would delete the email
            \item \textbf{Report} The participant indicated that they would report the email as spam
            \item \textbf{Unsubscribe} The participant indicated they would use the unsubscribe option within the email to prevent future emails
            \item \textbf{Block} The participant indicated they would block the sender
        \end{itemize}
\end{enumerate}

\subsection{Email Assessment, Training, Past Experience}

\begin{enumerate}
\itemsep 0.5em 

        \item \textbf{Dealing with Unsolicited Emails} Codes pertaining to cues, criteria or rules of thumb used to determine a legitimacy of an email.
        
        \begin{itemize}
            \item \textbf{Grammatical Cues} The participant expressed that they relied on cues such as grammatical inconsistencies, typos, misspellings, out-of-order symbols
            \item \textbf{Logical Cues} The participant expressed that they relied on logical cues such as the improbability of an email request
            \item \textbf{Elements in the Email} The participant expressed that they relied on cues in the email structure such as the subject, email sender, timestamp, and body without attachments

        \end{itemize}

        \item \textbf{Email Provider Sorting} Codes Pertaining to the participant’s opinion of how well their email provider is sorting their emails based on if they are spam or not.

        \begin{itemize}
            \item \textbf{Spam Going to Inbox} The participant reported that they had spam emails coming to their inbox frequently
            \item \textbf{Important Emails Going to Spam} The participant reported that they had an experience where an important email (job, school, requested materials, one-time password) went to their spam
        \end{itemize}

\end{enumerate}

\subsection{Usability and Improvements}

\begin{enumerate}
\itemsep 0.5em 

         
        \item \textbf{Email Usability Improvements} Codes pertaining to banner warning improvements
        \begin{itemize}
            \item \textbf{Severity/Risk Level Indicators} The participant recommends for the banner warnings to include severity/risk level indicators to better discriminate between various levels of threats and risk exposures based on the email type
            \item \textbf{Color Coding} The participant recommends for the banner warnings to be color coded to allow for better discrimination between various levels of threats and risk exposures based on the email type
            \item \textbf{Popup} The participant expressed that a warning be implemented as a pop-up that must be accepted before a suspicious link or attachment is opened
            \item \textbf{Message Content} The participant expressed that the content of the warning message could be more descriptive
            \item \textbf{Algorithm Control} The participant expressed that users should have a more active role in what is determined to be spam and the warnings received
         \end{itemize}
         
        \item \textbf{VR Usability} Codes pertaining to the ability to detect phishing and spam using a VR headset.
        \begin{itemize}
            \item \textbf{Misclicks} The participant expressed that it is too easy to click on phishing links or attachments
            \item \textbf{Distractions} The participant noted that there were too many distractions that they wouldn’t give the task the same level of focus
         \end{itemize}
         
        \item \textbf{VR Usability Improvements} Codes pertaining to suggestions to improve user’s abilities to detect spam and phishing when checking emails using a VR headset.
        \begin{itemize}
            \item \textbf{Screen Takeover} The participant expressed that the email browser/app should take over the full screen to reduce distractions
            \item \textbf{Navigation} The participant expressed that the navigation should be different when doing productivity tasks instead of gaming
        \end{itemize}
        
\end{enumerate}

\section{Debriefing} \label{sec:debrief}

Thank you for participating in our research on how users who are utilizing virtual reality headsets experience and utilize email warnings. This study aimed to examine whether people pay attention to warnings as a cue before they proceed to the website or not. So far, no research exists on how users are experiencing and utilizing emails warnings while utilizing virtual reality headsets. This is why we asked you to select and examine one email from your spam folder. \\

\noindent It was necessary for the researchers to withhold this information from you regarding the purpose of the study to ensure that your actions and answers to questions accurately reflected your cybersecurity hygiene, perceptions, and beliefs. Your participation in the study is important in helping researchers identify the best ways to address the accessibility of the warnings assigned by the email provided to the emails in the spam folder. Since we did not collect any personal information we would not able to remove your entry from the data bank of our research interviews enforced once you leave the research site.\\

\noindent The final results of this study will be published in a peer-reviewed journal. Your results will not be available individually and your participation will remain confidential. We do not keep, record, or collect any personal credentials. If you have any additional inquiries please contact \censor{Filipo Sharevski, Ph.D. at fsharevs@cdm.depaul.edu}. If you have questions about your rights as a research subject, you may contact \censor{Jessica Bloom} in the Office of Research Services at \censor{(312) 362-6168} or via email at \censor{jbloom8@depaul.edu}. You may also contact \censor{DePaul’s} Office of Research Services if your questions, concerns, or complaints are not being answered by the research team, you cannot reach the research team, or you want to talk to someone besides the research team. \\

%% file: main.bbl
\begin{thebibliography}{10}
\providecommand{\url}[1]{#1}
\csname url@samestyle\endcsname
\providecommand{\newblock}{\relax}
\providecommand{\bibinfo}[2]{#2}
\providecommand{\BIBentrySTDinterwordspacing}{\spaceskip=0pt\relax}
\providecommand{\BIBentryALTinterwordstretchfactor}{4}
\providecommand{\BIBentryALTinterwordspacing}{\spaceskip=\fontdimen2\font plus
\BIBentryALTinterwordstretchfactor\fontdimen3\font minus \fontdimen4\font\relax}
\providecommand{\BIBforeignlanguage}[2]{{%
\expandafter\ifx\csname l@#1\endcsname\relax
\typeout{** WARNING: IEEEtran.bst: No hyphenation pattern has been}%
\typeout{** loaded for the language `#1'. Using the pattern for}%
\typeout{** the default language instead.}%
\else
\language=\csname l@#1\endcsname
\fi
#2}}
\providecommand{\BIBdecl}{\relax}
\BIBdecl

\bibitem{Stephenson2022}
S.~Stephenson, B.~Pal, S.~Fan, E.~Fernandes, Y.~Zhao, and R.~Chatterjee, ``Sok: Authentication in augmented and virtual reality,'' in \emph{2022 IEEE Symposium on Security and Privacy (SP)}, 2022, pp. 267--284.

\bibitem{Freeman2022}
\BIBentryALTinterwordspacing
G.~Freeman, D.~Acena, N.~J. McNeese, and K.~Schulenberg, ``Working together apart through embodiment: Engaging in everyday collaborative activities in social virtual reality,'' \emph{Proc. ACM Hum.-Comput. Interact.}, vol.~6, no. GROUP, Jan. 2022. [Online]. Available: \url{https://doi.org/10.1145/3492836}
\BIBentrySTDinterwordspacing

\bibitem{Fu2022}
\BIBentryALTinterwordspacing
Y.~Fu, Y.~Hu, and V.~Sundstedt, ``A systematic literature review of virtual, augmented, and mixed reality game applications in healthcare,'' \emph{ACM Trans. Comput. Healthcare}, vol.~3, no.~2, Mar. 2022. [Online]. Available: \url{https://doi.org/10.1145/3472303}
\BIBentrySTDinterwordspacing

\bibitem{Drey2022}
\BIBentryALTinterwordspacing
T.~Drey, P.~Albus, S.~der Kinderen, M.~Milo, T.~Segschneider, L.~Chanzab, M.~Rietzler, T.~Seufert, and E.~Rukzio, ``Towards collaborative learning in virtual reality: A comparison of co-located symmetric and asymmetric pair-learning,'' in \emph{Proceedings of the 2022 CHI Conference on Human Factors in Computing Systems}, ser. CHI '22.\hskip 1em plus 0.5em minus 0.4em\relax New York, NY, USA: Association for Computing Machinery, 2022. [Online]. Available: \url{https://doi.org/10.1145/3491102.3517641}
\BIBentrySTDinterwordspacing

\bibitem{Ward2023}
\BIBentryALTinterwordspacing
A.~Ward, S.~Avula, H.-F. Cheng, S.~M. Sarwar, V.~Murdock, and E.~Agichtein, ``Searching for products in virtual reality: Understanding the impact of context and result presentation on user experience,'' in \emph{Proceedings of the 46th International ACM SIGIR Conference on Research and Development in Information Retrieval}, ser. SIGIR '23.\hskip 1em plus 0.5em minus 0.4em\relax New York, NY, USA: Association for Computing Machinery, 2023, p. 2359–2363. [Online]. Available: \url{https://doi.org/10.1145/3539618.3592057}
\BIBentrySTDinterwordspacing

\bibitem{Atopia2023}
\BIBentryALTinterwordspacing
Atopia, ``Who's really using vr these days? six data-driven insights into today's vr user demographic,'' Oct 2023. [Online]. Available: \url{https://medium.com/\@annabell\_37704/whos-really-using-vr-these-days-six-data-driven-insights-into-today-s-vr-user-demographic-422372a75c8c}
\BIBentrySTDinterwordspacing

\bibitem{Frankel2024}
\BIBentryALTinterwordspacing
D.~Frankel, ``While more than half of americans are interested in apple’s \$3,500 vision pro, 76\% say they have no intention of buying one,'' Apr 2024. [Online]. Available: \url{https://www.yahoo.com/tech/while-more-half-americans-interested-161909011.html}
\BIBentrySTDinterwordspacing

\bibitem{Gonzalez-Franco2024}
\BIBentryALTinterwordspacing
M.~Gonzalez-Franco and A.~Colaco, ``Guidelines for productivity in virtual reality,'' \emph{Interactions}, vol.~31, no.~3, p. 46–53, May 2024. [Online]. Available: \url{https://doi.org/10.1145/3658407}
\BIBentrySTDinterwordspacing

\bibitem{Apple2024}
\BIBentryALTinterwordspacing
A.~Inc, ``Apple vision pro.'' [Online]. Available: \url{https://www.apple.com/apple-vision-pro/}
\BIBentrySTDinterwordspacing

\bibitem{Hayden2020}
\BIBentryALTinterwordspacing
S.~Hayden, ``Facebook lays out the future of work and productivity on quest,'' Sep 2020. [Online]. Available: \url{https://www.roadtovr.com/facebook-future-work-productivity-quest-2/}
\BIBentrySTDinterwordspacing

\bibitem{Hart2024}
\BIBentryALTinterwordspacing
J.~Hart, ``I tried working a full day wearing apple’s vision pro. it’s the ultimate wfh device.'' [Online]. Available: \url{https://www.businessinsider.com/using-apple-vision-pro-for-the-day-productivity-2024-2}
\BIBentrySTDinterwordspacing

\bibitem{Axon2024}
\BIBentryALTinterwordspacing
S.~Axon, ``I worked exclusively in vision pro for a week-here’s how it went,'' Mar 2024. [Online]. Available: \url{https://arstechnica.com/gadgets/2024/03/i-worked-exclusively-in-vision-pro-for-a-week-heres-how-it-went/}
\BIBentrySTDinterwordspacing

\bibitem{Miller2024}
\BIBentryALTinterwordspacing
M.~Miller, ``I used the apple vision pro for my 8-hour work day, and it left me wanting more,'' Feb 2024. [Online]. Available: \url{https://www.zdnet.com/article/i-used-the-apple-vision-pro-for-my-8-hour-work-day-and-it-left-me-wanting-more/}
\BIBentrySTDinterwordspacing

\bibitem{Cheng2024}
R.~Cheng, N.~Wu, M.~Varvello, E.~Chai, S.~Chen, and B.~Han, ``A first look at immersive telepresence on apple vision pro,'' \emph{Proceedings of the 2024 ACM on Internet Measurement Conference}, p. 555–562, Nov 2024.

\bibitem{Guo2019}
J.~Guo, D.~Weng, Z.~Zhang, Y.~Liu, and Y.~Wang, ``Evaluation of maslows hierarchy of needs on long-term use of hmds – a case study of office environment,'' in \emph{2019 IEEE Conference on Virtual Reality and 3D User Interfaces (VR)}, 2019, pp. 948--949.

\bibitem{Grubert2018}
J.~Grubert, E.~Ofek, M.~Pahud, and P.~O. Kristensson, ``The office of the future: Virtual, portable, and global,'' \emph{IEEE Computer Graphics and Applications}, vol.~38, no.~6, pp. 125--133, 2018.

\bibitem{Makortoff2021}
\BIBentryALTinterwordspacing
K.~Makortoff, ``No more fomo: Top firms turn to vr to liven up meetings,'' Feb 2021. [Online]. Available: \url{https://www.theguardian.com/business/2021/feb/20/no-more-fomo-top-firms-turn-to-vr-to-liven-up-meetings}
\BIBentrySTDinterwordspacing

\bibitem{Mcgill2020}
\BIBentryALTinterwordspacing
M.~Mcgill, A.~Kehoe, E.~Freeman, and S.~Brewster, ``Expanding the bounds of seated virtual workspaces,'' \emph{ACM Trans. Comput.-Hum. Interact.}, vol.~27, no.~3, May 2020. [Online]. Available: \url{https://doi.org/10.1145/3380959}
\BIBentrySTDinterwordspacing

\bibitem{Burda2024}
\BIBentryALTinterwordspacing
P.~Burda, L.~Allodi, and N.~Zannone, ``Cognition in social engineering empirical research: A systematic literature review,'' \emph{ACM Trans. Comput.-Hum. Interact.}, vol.~31, no.~2, Jan. 2024. [Online]. Available: \url{https://doi.org/10.1145/3635149}
\BIBentrySTDinterwordspacing

\bibitem{Lain2022}
D.~Lain, K.~Kostiainen, and S.~Čapkun, ``Phishing in organizations: Findings from a large-scale and long-term study,'' in \emph{2022 IEEE Symposium on Security and Privacy (SP)}, 2022, pp. 842--859.

\bibitem{Distler2023}
\BIBentryALTinterwordspacing
V.~Distler, ``The influence of context on response to spear-phishing attacks: an in-situ deception study,'' in \emph{Proceedings of the 2023 CHI Conference on Human Factors in Computing Systems}, ser. CHI '23.\hskip 1em plus 0.5em minus 0.4em\relax New York, NY, USA: Association for Computing Machinery, 2023. [Online]. Available: \url{https://doi.org/10.1145/3544548.3581170}
\BIBentrySTDinterwordspacing

\bibitem{Petelka2019}
\BIBentryALTinterwordspacing
J.~Petelka, Y.~Zou, and F.~Schaub, ``Put your warning where your link is: Improving and evaluating email phishing warnings,'' in \emph{Proceedings of the 2019 CHI Conference on Human Factors in Computing Systems}, ser. CHI '19.\hskip 1em plus 0.5em minus 0.4em\relax New York, NY, USA: Association for Computing Machinery, 2019, p. 1–15. [Online]. Available: \url{https://doi.org/10.1145/3290605.3300748}
\BIBentrySTDinterwordspacing

\bibitem{Volkamer2017}
M.~Volkamer, K.~Renaud, B.~Reinheimer, and A.~Kunz, ``User experiences of torpedo: Tooltip-powered phishing email detection,'' \emph{Computers \& Security}, vol.~71, pp. 100--113, 2017.

\bibitem{Sharevski2024-usenix}
\BIBentryALTinterwordspacing
F.~Sharevski and A.~Zeidieh, ``Assessing suspicious emails with banner warnings among blind and {Low-Vision} users in realistic settings,'' in \emph{33rd USENIX Security Symposium (USENIX Security 24)}.\hskip 1em plus 0.5em minus 0.4em\relax Philadelphia, PA: USENIX Association, Aug. 2024, pp. 2083--2100. [Online]. Available: \url{https://www.usenix.org/conference/usenixsecurity24/presentation/sharevski}
\BIBentrySTDinterwordspacing

\bibitem{openphish}
{OpenPhish}, ``Openphish database,'' \url{https://openphish.com/phishing_database.html}.

\bibitem{Duzgun2022}
\BIBentryALTinterwordspacing
R.~D\"{u}zg\"{u}n, N.~Noah, P.~Mayer, S.~Das, and M.~Volkamer, ``Sok: A systematic literature review of knowledge-based authentication on augmented reality head-mounted displays,'' in \emph{Proceedings of the 17th International Conference on Availability, Reliability and Security}, ser. ARES '22.\hskip 1em plus 0.5em minus 0.4em\relax New York, NY, USA: Association for Computing Machinery, 2022. [Online]. Available: \url{https://doi.org/10.1145/3538969.3539011}
\BIBentrySTDinterwordspacing

\bibitem{Arafat2021}
A.~A. Arafat, Z.~Guo, and A.~Awad, ``Vr-spy: A side-channel attack on virtual key-logging in vr headsets,'' in \emph{2021 IEEE Virtual Reality and 3D User Interfaces (VR)}, 2021, pp. 564--572.

\bibitem{Gopal2023}
\BIBentryALTinterwordspacing
S.~R.~K. Gopal, D.~Shukla, J.~D. Wheelock, and N.~Saxena, ``Hidden reality: Caution, your hand gesture inputs in the immersive virtual world are visible to all!'' in \emph{32nd USENIX Security Symposium (USENIX Security 23)}.\hskip 1em plus 0.5em minus 0.4em\relax Anaheim, CA: USENIX Association, Aug. 2023, pp. 859--876. [Online]. Available: \url{https://www.usenix.org/conference/usenixsecurity23/presentation/gopal}
\BIBentrySTDinterwordspacing

\bibitem{Khalili2024}
H.~Khalili, A.~Chen, T.~Papaiakovou, T.~Jacques, H.-J. Chien, C.~Liu, A.~Ding, A.~Hass, S.~Zonouz, and N.~Sehatbakhsh, ``Virtual keymysteries unveiled: Detecting keystrokes in vr with external side-channels,'' in \emph{2024 IEEE Security and Privacy Workshops (SPW)}, 2024, pp. 260--266.

\bibitem{luo2024eavesdropping}
S.~Luo, A.~Nguyen, H.~Farooq, K.~Sun, and Z.~Yan, ``Eavesdropping on controller acoustic emanation for keystroke inference attack in virtual reality,'' in \emph{The Network and Distributed System Security Symposium (NDSS)}, 2024.

\bibitem{Slocum2023}
\BIBentryALTinterwordspacing
C.~Slocum, Y.~Zhang, N.~Abu-Ghazaleh, and J.~Chen, ``Going through the motions: {AR/VR} keylogging from user head motions,'' in \emph{32nd USENIX Security Symposium (USENIX Security 23)}.\hskip 1em plus 0.5em minus 0.4em\relax Anaheim, CA: USENIX Association, Aug. 2023, pp. 159--174. [Online]. Available: \url{https://www.usenix.org/conference/usenixsecurity23/presentation/slocum}
\BIBentrySTDinterwordspacing

\bibitem{wang2024gazeploit}
H.~Wang, Z.~Zhan, H.~Shan, S.~Dai, M.~Panoff, and S.~Wang, ``Gazeploit: Remote keystroke inference attack by gaze estimation from avatar views in vr/mr devices,'' \emph{arXiv preprint arXiv:2409.08122}, 2024.

\bibitem{yang2024inception}
Z.~Yang, C.~Y. Li, A.~Bhalla, B.~Y. Zhao, and H.~Zheng, ``Inception attacks: Immersive hijacking in virtual reality systems,'' \emph{arXiv preprint arXiv:2403.05721}, 2024.

\bibitem{Verizon.2022}
Verizon, ``{Data Breach Investigations Report 2008 -- 2022},'' {Verizon}, Tech. Rep., 2023, \url{https://www.verizon.com/business/resources/Tabb/reports/2022-data-breach-investigations-report-dbir.pdf}.

\bibitem{Verizon.2023}
------, ``{Data Breach Investigations Report 2023},'' {Verizon}, Tech. Rep., 2023, \url{https://www.verizon.com/business/resources/Tabb/reports/2023-data-breach-investigations-report-dbir.pdf}.

\bibitem{Kumaran2022}
\BIBentryALTinterwordspacing
N.~Kumaran, ``An overview of gmail’s spam filters | google workspace blog,'' May 2022. [Online]. Available: \url{https://workspace.google.com/blog/identity-and-security/an-overview-of-gmails-spam-filters?hl=en}
\BIBentrySTDinterwordspacing

\bibitem{Microsoft2024}
\BIBentryALTinterwordspacing
Microsoft, ``Microsoft,'' Apr 2024. [Online]. Available: \url{https://support.microsoft.com/en-us/office/overview-of-the-junk-email-filter-5ae3ea8e-cf41-4fa0-b02a-3b96e21de089}
\BIBentrySTDinterwordspacing

\bibitem{Franz2021}
\BIBentryALTinterwordspacing
A.~Franz, V.~Zimmermann, G.~Albrecht, K.~Hartwig, C.~Reuter, A.~Benlian, and J.~Vogt, ``{SoK}: Still plenty of phish in the sea {\textemdash} a taxonomy of {User-Oriented} phishing interventions and avenues for future research,'' in \emph{Seventeenth Symposium on Usable Privacy and Security (SOUPS 2021)}.\hskip 1em plus 0.5em minus 0.4em\relax USENIX Association, Aug. 2021, pp. 339--358. [Online]. Available: \url{https://www.usenix.org/conference/soups2021/presentation/franz}
\BIBentrySTDinterwordspacing

\bibitem{GooglePhishing}
{Google}, ``Advanced phishing and malware protection,'' 2023, \url{https://support.google.com/a/answer/9157861?hl=en}.

\bibitem{OutlookPhishing}
{Microsoft}, ``Overview of the junk email filter,'' 2023, \url{https://support.microsoft.com/en-us/office/overview-of-the-junk-email-filter-5ae3ea8e-cf41-4fa0-b02a-3b96e21de089}.

\bibitem{Risher-Miller.2017}
M.~Risher and A.~Miller, ``Fighting phishing with smarter protections,'' 2017, \url{https://blog.google/technology/safety-security/fighting-phishing-smarter-protections/}.

\bibitem{Wash2020}
\BIBentryALTinterwordspacing
R.~Wash, ``How experts detect phishing scam emails,'' \emph{Proc. ACM Hum.-Comput. Interact.}, vol.~4, no. CSCW2, oct 2020. [Online]. Available: \url{https://doi.org/10.1145/3415231}
\BIBentrySTDinterwordspacing

\bibitem{Mossano2023}
\BIBentryALTinterwordspacing
M.~Mossano, O.~Kulyk, B.~M. Berens, E.~M. H\"{a}u\ss{}ler, and M.~Volkamer, ``Influence of url formatting on users' phishing url detection,'' in \emph{Proceedings of the 2023 European Symposium on Usable Security}, ser. EuroUSEC '23.\hskip 1em plus 0.5em minus 0.4em\relax New York, NY, USA: Association for Computing Machinery, 2023, p. 318–333. [Online]. Available: \url{https://doi.org/10.1145/3617072.3617111}
\BIBentrySTDinterwordspacing

\bibitem{milgram1994taxonomy}
P.~Milgram and F.~Kishino, ``A taxonomy of mixed reality visual displays,'' \emph{IEICE TRANSACTIONS on Information and Systems}, vol.~77, no.~12, pp. 1321--1329, 1994.

\bibitem{Kanaoka2024}
A.~Kanaoka and T.~Isohara, ``Enhancing smishing detection in ar environments: Cross-device solutions for seamless reality,'' in \emph{2024 IEEE Conference on Virtual Reality and 3D User Interfaces Abstracts and Workshops (VRW)}, 2024, pp. 565--572.

\bibitem{JansenFischbach2020}
\BIBentryALTinterwordspacing
P.~Jansen and F.~Fischbach, ``The social engineer: An immersive virtual reality educational game to raise social engineering awareness,'' in \emph{Extended Abstracts of the 2020 Annual Symposium on Computer-Human Interaction in Play}, ser. CHI PLAY '20.\hskip 1em plus 0.5em minus 0.4em\relax New York, NY, USA: Association for Computing Machinery, 2020, p. 59–63. [Online]. Available: \url{https://doi.org/10.1145/3383668.3419917}
\BIBentrySTDinterwordspacing

\bibitem{Bakker2024}
\BIBentryALTinterwordspacing
S.~{Bakker}, ``Immersive virtual reality and cybersecurity: Combatting social engineering in a healthcare context,'' July 2024. [Online]. Available: \url{http://essay.utwente.nl/100811/}
\BIBentrySTDinterwordspacing

\bibitem{Wogalter2002}
\BIBentryALTinterwordspacing
M.~S. Wogalter, V.~C. Conzola, and T.~L. Smith-Jackson, ``Research-based guidelines for warning design and evaluation,'' \emph{Applied Ergonomics}, vol.~33, no.~3, pp. 219--230, 2002. [Online]. Available: \url{https://www.sciencedirect.com/science/article/pii/S0003687002000091}
\BIBentrySTDinterwordspacing

\bibitem{Vigderman2024}
\BIBentryALTinterwordspacing
Vigderman, ``Virtual reality awareness and adoption report,'' January 2024. [Online]. Available: \url{https://www.security.org/digital-security/virtual-reality-annual-report/}
\BIBentrySTDinterwordspacing

\bibitem{Petrov2024}
\BIBentryALTinterwordspacing
C.~Petrov, ``45 virtual reality statistics that rock the market in 2024,'' Jan 2024. [Online]. Available: \url{https://techjury.net/blog/virtual-reality-statistics/gref}
\BIBentrySTDinterwordspacing

\bibitem{cisa-phishing}
{Cybersecurity and Infrastrucure Security Agency (CISA)}, ``Avoiding social engineering and phishing attacks,'' 2021, \url{https://www.cisa.gov/news-events/news/avoiding-social-engineering-and-phishing-attacks}.

\bibitem{clarke2015thematic}
V.~Clarke, V.~Braun, and N.~Hayfield, ``Thematic analysis,'' \emph{Qualitative psychology: A practical guide to research methods}, vol.~3, pp. 222--248, 2015.

\bibitem{ortloff2023}
\BIBentryALTinterwordspacing
A.-M. Ortloff, M.~Fassl, A.~Ponticello, F.~Martius, A.~Mertens, K.~Krombholz, and M.~Smith, ``Different researchers, different results? analyzing the influence of researcher experience and data type during qualitative analysis of an interview and survey study on security advice,'' in \emph{Proceedings of the 2023 CHI Conference on Human Factors in Computing Systems}, ser. CHI '23.\hskip 1em plus 0.5em minus 0.4em\relax New York, NY, USA: Association for Computing Machinery, 2023. [Online]. Available: \url{https://doi.org/10.1145/3544548.3580766}
\BIBentrySTDinterwordspacing

\bibitem{fereday2006demonstrating}
J.~Fereday and E.~Muir-Cochrane, ``Demonstrating rigor using thematic analysis: A hybrid approach of inductive and deductive coding and theme development,'' \emph{International journal of qualitative methods}, vol.~5, no.~1, pp. 80--92, 2006.

\bibitem{Bryant2023}
\BIBentryALTinterwordspacing
C.~Bryant, Z.~Yuan, M.~R. Qorib, H.~Cao, H.~T. Ng, and T.~Briscoe, ``{Grammatical Error Correction: A Survey of the State of the Art},'' \emph{Computational Linguistics}, vol.~49, no.~3, pp. 643--701, 09 2023. [Online]. Available: \url{https://doi.org/10.1162/coli\_a\_00478}
\BIBentrySTDinterwordspacing

\bibitem{glennon2018use}
C.~Glennon, ``Use of virtual reality to distract from pain and anxiety,'' \emph{Number 4/July 2018}, vol.~45, no.~4, pp. 545--552, 2018.

\bibitem{Petelka2025}
\BIBentryALTinterwordspacing
J.~Petelka, B.~Berens, C.~Sugatan, M.~Volkamer, and F.~Schaub, ``{ Restricting the Link: Effects of Focused Attention and Time Delay on Phishing Warning Effectiveness },'' in \emph{2025 IEEE Symposium on Security and Privacy (SP)}.\hskip 1em plus 0.5em minus 0.4em\relax Los Alamitos, CA, USA: IEEE Computer Society, May 2025, pp. 7--7. [Online]. Available: \url{https://doi.ieeecomputersociety.org/10.1109/SP61157.2025.00007}
\BIBentrySTDinterwordspacing

\bibitem{Berens2024}
\BIBentryALTinterwordspacing
B.~M. Berens, F.~Schaub, M.~Mossano, and M.~Volkamer, ``Better together: The interplay between a phishing awareness video and a link-centric phishing support tool,'' in \emph{Proceedings of the 2024 CHI Conference on Human Factors in Computing Systems}, ser. CHI '24.\hskip 1em plus 0.5em minus 0.4em\relax New York, NY, USA: Association for Computing Machinery, 2024. [Online]. Available: \url{https://doi.org/10.1145/3613904.3642843}
\BIBentrySTDinterwordspacing

\bibitem{schoni2024you}
L.~Sch{\"o}ni, V.~Carles, M.~Strohmeier, P.~Mayer, and V.~Zimmermann, ``You know what?-evaluation of a personalised phishing training based on users’ phishing knowledge and detection skills,'' in \emph{The 2024 European Symposium on Usable Security}, 2024.

\bibitem{Ho2025}
\BIBentryALTinterwordspacing
G.~Ho, A.~Mirian, E.~Luo, K.~Tong, E.~Lee, L.~Liu, C.~A. Longhurst, C.~Dameff, S.~Savage, and G.~M. Voelker, ``{ Understanding the Efficacy of Phishing Training in Practice },'' in \emph{2025 IEEE Symposium on Security and Privacy (SP)}.\hskip 1em plus 0.5em minus 0.4em\relax Los Alamitos, CA, USA: IEEE Computer Society, May 2025, pp. 76--76. [Online]. Available: \url{https://doi.ieeecomputersociety.org/10.1109/SP61157.2025.00076}
\BIBentrySTDinterwordspacing

\bibitem{Saffo2021}
\BIBentryALTinterwordspacing
D.~Saffo, S.~Di~Bartolomeo, C.~Yildirim, and C.~Dunne, ``Remote and collaborative virtual reality experiments via social vr platforms,'' in \emph{Proceedings of the 2021 CHI Conference on Human Factors in Computing Systems}, ser. CHI '21.\hskip 1em plus 0.5em minus 0.4em\relax New York, NY, USA: Association for Computing Machinery, 2021. [Online]. Available: \url{https://doi.org/10.1145/3411764.3445426}
\BIBentrySTDinterwordspacing

\bibitem{Zreality2024}
\BIBentryALTinterwordspacing
Zreality, ``The market for virtual reality headsets in 2024: Meta dominates, apple vision pro enters the market,'' Feb 2024. [Online]. Available: \url{https://www.zreality.com/the-market-for-virtual-reality-headsets-in-2024-meta-dominates-apple-vision-pro-enters-the-market/}
\BIBentrySTDinterwordspacing

\bibitem{Greenwald2024}
\BIBentryALTinterwordspacing
W.~Greenwald, ``The best vr headsets for 2024,'' Nov 2024. [Online]. Available: \url{https://www.pcmag.com/picks/the-best-vr-headsets}
\BIBentrySTDinterwordspacing

\bibitem{Mott2019}
M.~Mott, E.~Cutrell, M.~Gonzalez~Franco, C.~Holz, E.~Ofek, R.~Stoakley, and M.~Ringel~Morris, ``Accessible by design: An opportunity for virtual reality,'' in \emph{2019 IEEE International Symposium on Mixed and Augmented Reality Adjunct (ISMAR-Adjunct)}, 2019, pp. 451--454.

\bibitem{Yu2023}
Y.~Yu, S.~Ashok, S.~Kaushik, Y.~Wang, and G.~Wang, ``Design and evaluation of inclusive email security indicators for people with visual impairments,'' in \emph{2023 IEEE Symposium on Security and Privacy (SP)}, 2023, pp. 2885--2902.

\bibitem{Abeele2021}
\BIBentryALTinterwordspacing
V.~V. Abeele, B.~Schraepen, H.~Huygelier, C.~Gillebert, K.~Gerling, and R.~Van~Ee, ``Immersive virtual reality for older adults: Empirically grounded design guidelines,'' \emph{ACM Trans. Access. Comput.}, vol.~14, no.~3, Aug. 2021. [Online]. Available: \url{https://doi.org/10.1145/3470743}
\BIBentrySTDinterwordspacing

\end{thebibliography}
